\documentclass{article}
\usepackage[utf8]{inputenc}
\usepackage[T1]{fontenc}
\usepackage{authblk}
\usepackage[pdftex]{graphicx}
\newcommand{\etal}{\emph{et al.}\xspace}

\newcommand{\qmmm}{QM/MM\xspace}
\newcommand{\qm}{QM\xspace}
\newcommand{\mm}{MM\xspace}

\newcommand{\figrff}[1]{Figure~\ref{#1}}

\newcommand{\amoeba}{{\sc AMOEBA}\xspace}
\newcommand{\gaussian}{{\sc GAUSSIAN}\xspace}
\newcommand{\lichem}{{\sc LICHEM}\xspace}
\newcommand{\tinker}{{\sc Tinker}\xspace}
\newcommand{\tinkerhp}{{\sc Tinker-HP}\xspace}

\usepackage{amsmath}
\usepackage{amsfonts}
\usepackage{amssymb}
\usepackage{xspace}
\usepackage{subfig}
\usepackage{float}

\title{Development and Application of QM/MM Methods with Advanced Polarizable Potentials}

\author[1]{Jorge Nochebuena}
\author[1]{Sehr Naseem-Khan}
\author[1*]{Andrés Cisneros}

\affil{Department of Chemistry, University of North Texas, Denton, Texas, 76201, USA}
\affil[*]{Corresponding author: G. Andrés Cisneros (andres@unt.edu)}

\begin{document}
\maketitle
\begin{abstract}
Quantum Mechanics/Molecular Mechanics (\qmmm) simulations are a popular approach to study various features of large systems.
A common application of \qmmm calculations is in the investigation of reaction mechanisms in condensed--phase and biological systems. 
The combination of QM and MM methods to represent a system gives rise to several challenges that need to be
addressed.
The increase in computational speed has allowed
the expanded use of more complicated and accurate methods
for both QM and MM simulations. 
Here, we review some approaches that address several common challenges encountered in \qmmm simulations with
advanced polarizable potentials, from methods to account
for boundary across covalent bonds and long--range
effects, to polarization and advanced embedding
potentials.
\end{abstract}

\section{Introduction}
\label{sec:intro}
The simulation of reaction processes in complex biological systems such as enzymes, or condensed--phase systems, requires the description of changes in electronic and nuclear degrees of freedom.
Although quantum methods can adequately describe electronic effects, the modeling of the systems of interest is limited by the size and the time scale that can be carried out in reasonable computational times.
By contrast, molecular mechanics methods allow the study of systems with millions atoms but are limited to the type of system for which the force field was designed, and most force fields used for these simulations are non--reactive \cite{Xiao2018}.

\begin{figure}[t!]
    \centering
    \includegraphics[width=1\textwidth]{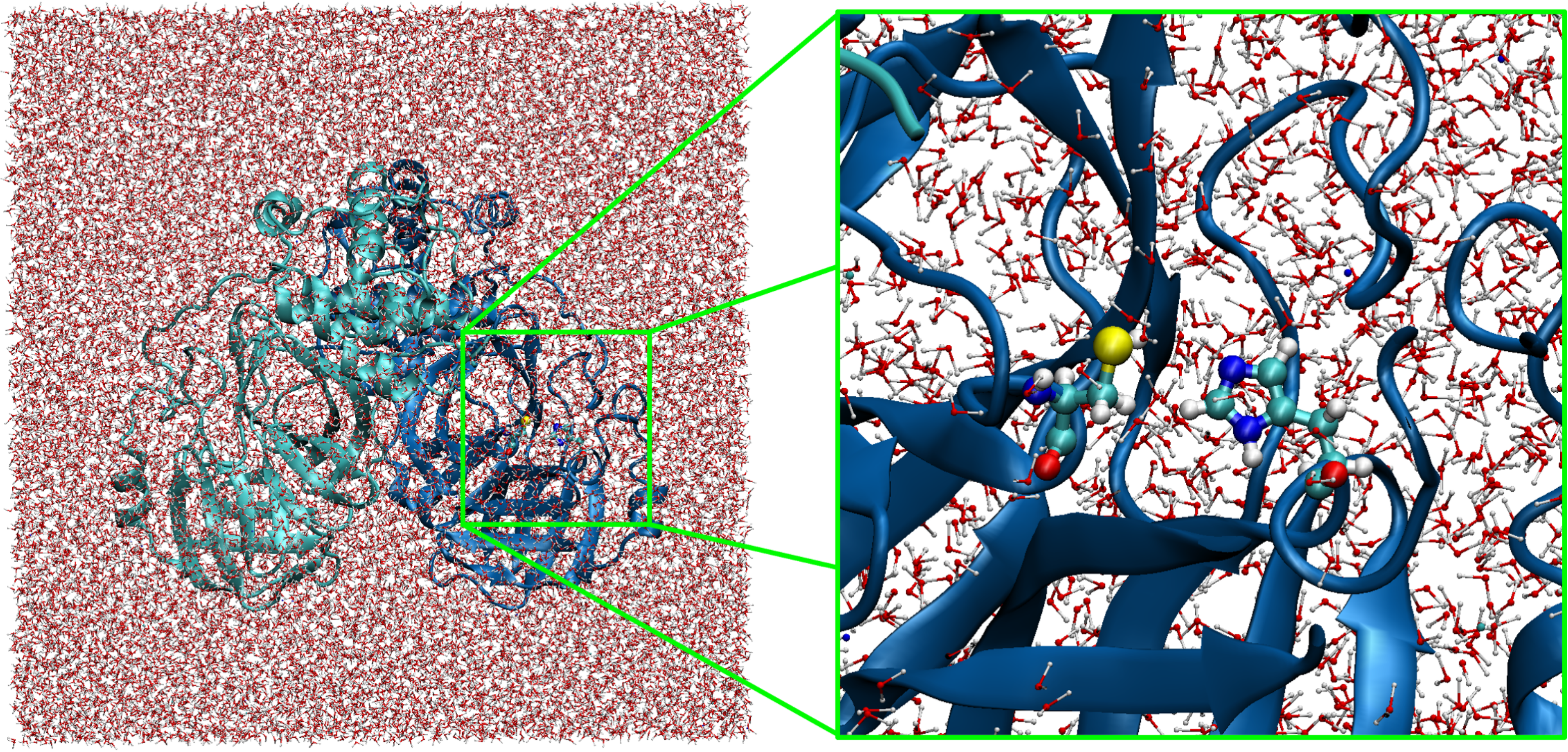}
    \caption{Main protease of SARS-CoV2 in a box of water \cite{Komatsu2020}
    The whole protein (left) and a close-up in the active site (right) are shown. Histidine 41 and cysteine 145 amino acid residues are represented by balls and cylinders. PDB: 6LU7 \cite{Jin2020}.}
    \label{fig:proton_transfer}
\end{figure}

Quantum mechanics/molecular mechanics (QM/MM) methods emerged in the mid-1970s as a proposal to study enzymatic reactions \cite{Warshel1976}.
Since then, QM/MM methods have been widely applied to study various chemical and biochemical systems \cite{Lin2006,Senn2009,VanderKamp2013,Brunk2015,Duarte2015,Ahmadi2018,Magalhaes2020}.
QM/MM methods aim to leverage the advantages of both quantum
and classical approaches by combining two levels of theory.
The quantum model focuses on the region of the system where the reactive
processes of interest take place, while the classical model 
describes the rest of the system.
For example, if the process involves a reaction in the active site of an enzyme (see Figure \ref{fig:proton_transfer}), the residues and any other molecules (or fragments) involved in the reaction will be assigned as the QM subsystem, and the rest of the enzyme, solvent, counterions, etc. will be represented by the MM force field.

There are a wide variety of combinations for the treatment of the QM and MM subsystem.
The QM subsystem may be represented by empirical valence bond (EVB) \cite{Warshel1980,Aqvist1993,Kamerlin2010}, semi--empirical \cite{Govender2014,Ojeda-May2017} or \emph{ab initio} Hamiltonians \cite{Strajbl2002,Intharathep2006,Hu2009,Fang2013}.
For the MM environment many \qmmm simulations employ non--polarizable point--charge force fields (npFFs) \cite{Brooks1983,Jorgensen1996,Case2005}.
More recently, increase in computational
speed and algorithmic developments have allowed the use
of more accurate potentials which may include explicit
representation of electronic polarization, more elaborate
electrostatics, and in some cases, explicit inclusion
of other quantum effects \cite{Hagras2018,Vazquez-Montelongo2018,Giovannini2019,Giovannini2019b}.

The coupling of QM and MM methods to represent a system
gives rise to several challenges such as how to couple
these two levels of theory across covalent boundaries,
how to treat long--range effects in the context of QM
subsystems, coupling of the quantum and classical 
Hamiltonians and how to treat the explicit couplings,
among others. In this review, we present approaches
that tackle several of these issues.

This article is organized as follows.
Section \ref{sec:background} provides an overview of the \qmmm approach.
It describes what force fields are, the different types of embedding, as well as some methods to include polarization.
Section \ref{sec:qmmm-boundary} describes the challenges involved in separating the QM and MM regions across covalent bonds, and solutions that have been proposed.
The differences between the link atom, double link atom, as well as LSCF and GHO methods will be described.
In subsection \ref{sec:qmmm-elec} the role played by long-range interactions and methods that have been developed to include them efficiently will be described.
In section \ref{sec:qmmm-adv-pot} we will describe some recent development for \qmmm simulations using advanced polarizable force fields.
Finally, in section \ref{sec:lichem} we will show some of the methods implemented in LICHEM, a code developed in the group to perform \qmmm simulations, followed by concluding remarks.

\section{Classical MM environments in \qmmm}
\label{sec:background}
Many \qmmm simulations rely on the combination of QM methods
with the use of potential-energy functions, called force fields \cite{Notman2013}.
Many force fields approximate the energy and forces of the system by separating the contributions into bonded and non--bonded interactions.
In several cases, the non--bonded interactions are
approximated by Coulomb and Van der Waals interactions.
Some examples are: AMBER \cite{Case2005}, CHARMM \cite{Brooks1983}, GROMOS \cite{Scott1999}, MMFF \cite{Halgren1999}, MM3 \cite{Allinger1989}, MM4 \cite{Allinger1996}, OPLS \cite{Jorgensen1996} and UFF \cite{Rappe1992}.

A common expression for fixed--charge, non--polarizable
force fields is show in  Eq. \eqref{mm_energy}
\begin{align}
    E_\textnormal{int} = & \sum_\textnormal{bonds} K_b (b-b_0)^2 + \sum_\textnormal{angles} K_\theta (\theta-\theta_0)^2 + \sum_\textnormal{dihedrals} K_\phi [1+\cos(n\phi+\delta)] \notag \\
    & + \sum_{i<j} \frac{q_i q_j}{r_{ij}} + \sum_{i<j} 4\varepsilon_{ij} \left[ \left( \frac{\sigma_{ij}}{r_{ij}} \right)^{12} - \left( \frac{\sigma_{ij}}{r_{ij}}\right)^{6}\right],
    \label{mm_energy}
\end{align} 
where the first three terms are the bonded contributions and the last two represent non-bonded interactions. $K_b$, $K_\theta$ and $K_\phi$ are force constants for bonds, angles and dihedral terms; $b_0$ and $\theta_0$ are equilibrium values for bond length and angles between atoms; $n$ is the dihedral multiplicity; $\delta$ is the dihedral angle phase; $q$ are
partial (generally atomic) charges; $\varepsilon$ are the well depths, 
and $\sigma$ are the van der Waals radii.
In the above equation, and throughout the rest of 
the review all equations use atomic units.

Given the functional form of the potential, the total energy for a \qmmm system can be separated in 3 contributions:
the contribution from the QM subsystem, the contribution from the MM 
environment, and the interaction between both regions.
The first two contributions are straightforward to evaluate.
In principle, any combination of QM and MM methods can be chosen.
The interaction between the QM and MM regions poses some challenges as mentioned above.
The combination of the levels of theory can be achieved by two main approaches termed the {\em subtractive} and 
{\em additive} coupling schemes.

In the subtractive method the \qmmm energy is obtained by:
\begin{equation}
    E_\textnormal{\qmmm}^\textnormal{sub} =
    E_\textnormal{\mm} (\textnormal{\mm} + \textnormal{\qm}) + E_\textnormal{\qm} (\textnormal{\qm}) - E_\textnormal{\mm} (\textnormal{\qm}),
    \label{substractive_method}
\end{equation}
that is, first the energy of the complete system (QM and MM regions) is evaluated at the MM level, then the energy of the QM region calculated in the QM level is added, and finally, the energy of the QM region evaluated at the MM level is subtracted.
This method is straightforward but has the disadvantage that it requires several calculations at different levels of theory of the same set of atoms.

In the additive scheme, the QM/MM energy is calculated by:
\begin{equation}
    E_\textnormal{\qmmm}^\textnormal{add} =
    E_\textnormal{\qm} (\textnormal{\qm}) + V_\textnormal{\mm} (\textnormal{\mm}) + E_\textnormal{\qmmm} (\textnormal{\qm} + \textnormal{\mm}),
    \label{additive_method}
\end{equation}
that is, the energy \qmmm is the sum of the contributions QM and MM regions evaluated at its own level plus the contribution due to coupling.
The complexity of the \qmmm methods depends on how the coupling term is expressed.

The last term in Eq. \eqref{additive_method} can be further subdivided depending on the individual terms
of the classical force field.
If the force field uses a functional form as in Eq. \eqref{mm_energy}, the QM/MM interaction can be divided as:
\begin{equation}
     E_\textnormal{\qmmm}^\textnormal{total} = 
     E_\textnormal{\qmmm}^\textnormal{bonds} +
     E_\textnormal{\qmmm}^\textnormal{angles} +
     E_\textnormal{\qmmm}^\textnormal{torsions} +
     E_\textnormal{\qmmm}^\textnormal{Coulomb} +
     E_\textnormal{\qmmm}^\textnormal{VDW}.
    \label{additive_method_sub}
\end{equation}

In both substractive and attractive schemes, it is possible to further subdivide the approaches depending on how each of the terms in the \qmmm interaction is calculated.
In the {\em mechanical embedding} approach, the interaction between the QM and MM regions are handled at the force field level.
The QM subsystem is thus kept in place by MM interactions.
Similarly to Eq. \eqref{mm_energy}, all bonded terms involving QM atoms are represented with the respective MM functions, and all non--bonded interactions involving QM and MM atoms are represented by point charges and Lennard--Jones potential energy terms.
In this approach, the calculation of the QM system is performed isolated from the MM region, and therefore, the wave function is not explicitly polarized by the external field from the MM environment.

In the {\em electrostatic embedding} scheme, the electrostatic interactions between the two regions,
$E_\textnormal{\qmmm}^\textnormal{Coulomb}$, are included in the QM calculation.
In this case, the MM atoms polarize the electrons in the QM subsystem via an effective Hamiltonian,
\begin{equation}
    H_{\textnormal{\textnormal{eff}}} =
    H_{\textnormal{\textnormal{\qm}}} - \sum_j^M \frac{q_j}{|\textbf{r}_i-\textbf{R}_j|},
    \label{electrostatic_embedding}
\end{equation}
where $H_{\textnormal{\qm}}$ is the electronic Hamiltonian for the QM
subsystem, $M$ is the number of MM atoms that have a partial charge $q_j$, and $\textbf{r}_i$ and $\textbf{R}_j$ are the positions of electron $i$ and MM atom $j$.
In this approach, the QM subsystem interacts with the MM region as a 
set of external point charges (or higher order multipoles).
However, mutual polarization effects or other non-electrostatic interactions are not taken into account.

\subsection{Polarizable QM/MM}

The force fields mentioned so far are extremely efficient and provide a convenient approach to investigate myriad
systems. These force fields in some cases can
exhibit limited accuracy in the reproduction of certain regions of the potential energy surface.
One major reason for the reduced accuracy involves the approximations employed to represent the bonded and non--bonded interactions.
In many cases, the non--bonded interactions are represented by two separate terms assuming pair--wise additivity.
These terms aim to represent the underlying physical interactions, however there are several approximations that are employed, which lead to the neglect of some physical interactions.
In general, the electrostatic interactions between particles are  approximated by fixed atomic partial charges. Dispersion and repulsion effects are approximated by a Lennard--Jones (or similar) function \cite{Munoz-Munoz2015,VanVleet2016,Boulanger2018}.

The use of fixed atomic point charges may be unable to capture or accurately represent several important effects such as charge density anisotropy, and charge penetration.
In addition, the approximation of the charge density by fixed point charges neglects certain many-body effects such as explicitly
accounting for electronic polarization or charge--transfer.
Polarization effects can be included by explicitly taking into account the effect of the electric field on the charge distribution of the constituent molecules. Several polarizable force fields have been developed including
AMOEBA \cite{Ponder2010}, CHARMM \cite{Lopes2009b}, EFP  \cite{Gordon2012}, GEM \cite{Cisneros2005,Piquemal2006}, GROMOS \cite{Geerke2007}, NEMO \cite{Holt2009} and SIBFA \cite{Gresh2007}, which employ one
of the methods below to describe polarization effects.


 
In the case where the force field has an explicit term
to represent the electronic polarization of the classical environment, it is necessary to account for the interaction between this effect in the MM and the QM subsystem.
In this case, the {\em polarization embedding} scheme is such that the QM and MM regions are mutually polarized.
Thus, not only the QM region is polarized by the QM charges, but also the MM region responds to the chemical environment in the QM region.
Three basic types of polarizable models have been developed: fluctuating charge model \cite{Lipparini2011}, Drude oscillator model \cite{Boulanger2012,Boulanger2012}, and induced dipole model \cite{Loco2016,Bondanza2020}.

The {\em fluctuating charge} (FQ) model is one of the most straightforward polarizable models.
It is based on the principle of electronegativity equalization.
That is, a charge oscillates between the atoms until the electronegativities of the atoms are equalized.
This method allows the modification of the value of the partial charges in response to the electric field, thus altering molecular polarization.
This can be done by coupling the charges to the environment using electronegativity or chemical potential equalization schemes.
FQ method uses a Taylor expansion truncated up to second-order terms of the atomic chemical potential to express the energy required to create a charge, Q, 
\begin{equation}
    E(Q) = E^0 + \left( \frac{\partial E}{\partial Q} \right)_0 Q + \frac{1}{2} \left( \frac{\partial^2 E}{\partial Q^2} \right)_0 Q^2
    \label{fluctuation_charge}
\end{equation}
The energies required to create +1 and -1 charges on an atom, obtained from Eq. \eqref{fluctuation_charge} 
\begin{align}
    E(+1) & =
    E^0 + \left( \frac{\partial E}{\partial Q} \right)_0 +
    \frac{1}{2} \left( \frac{\partial^2E}{\partial Q^2} \right)_0 \\
    E(-1) & =
    E^0 - \left( \frac{\partial E}{\partial Q} \right)_0 +
    \frac{1}{2} \left( \frac{\partial^2E}{\partial Q^2} \right)_0
\end{align}
using the definition of ionization potential as $IP=E(+1)-E(0)$ and electron affinity as $EA=E(0)-E(-1)$ we obtain:
\begin{align}
    \left( \frac{\partial E}{\partial Q} \right)_0 & = \frac{1}{2} (IP + EA) = \chi^0 \\
    \left( \frac{\partial^2 E}{\partial Q^2} \right)_0 & = IP - EA = J^0 = 2 \eta^0
\end{align}
where the $\chi^0$ is the electronegativity  and $\eta^0$ is the chemical 
hardness.  Equation \eqref{fluctuation_charge} can be rewritten as:
\begin{equation}
    E(Q) = E^0 +\chi^0Q + \eta^0 Q^2
\end{equation}
The previous equation can be generalized for a set of $N$ atoms
\begin{equation}
    E(Q_1...Q_N) = \sum_i \left( E_{i0} + \chi_i^0 Q_i + n_{ii}^0 \right) + \sum_{i,j>i} 2\eta_{ij}Q_iQ_j
\end{equation}
The optimal charge distribution is achieved by minimizing the energy with respect to the charges on each atom:
\begin{equation}
    \frac{\partial E}{\partial Q_{i}} = 0,
    \hspace{0.7cm}
    i=1,N_i
\end{equation}
 The constraints used in solving the previous equation include total charge conservation:
\begin{equation}
    Q_{\textnormal{tot}} =
    \sum_{i=1}^N Q_i
\end{equation}
and electronegativity equalization at convergence:
\begin{equation}
    \chi_1 = \chi_2 ... =\chi_N
\end{equation}

The {\em Drude oscillator} model incorporates electronic polarizability by representing an atom as a two-particle system: a charged core with charge $q_0$ and a charged shell, also called a Drude particle, with charge $q_D$.
The core and shell are linked by a harmonic spring.
In the absence of an electric field, the Drude particle is located at the atomic core. Otherwise, the Drude particle will be at a distance $d$ from the atomic core in the presence of an electric field:
\begin{equation}
    d = \frac{q_D \textnormal{E}}{k_D}
\end{equation}
The atomic induced dipole then is treated as:
\begin{equation}
    \mu = q_D d = q_D \times \frac{q_D E}{k_D} = \frac{q_D^2 E}{k_D}
\end{equation}
The atomic polarizability, $\alpha$, is related to the force constant $k$ of the harmonic spring connecting the core and shell and is determined by
\begin{equation}
    \alpha = \frac{q_D^2}{k_D}
\end{equation}
In principle, the Drude oscillator, $q_D$, the force constant of the harmonic spring, $k_D$, or both can be tuned to achieve an appropriate polarization response.

In the {\em induced dipole} model the polarization response is represented by a set of inducible dipoles that arise due to the external permanent and induced electric field. 
The induced dipole moment at the site $i$ is:
\begin{equation}
    \mu_i =
    \alpha_i \left[ E_i - \sum_{j \neq i}^N T_{ij} \mu_j \right]
\end{equation}
where $T_{ij}$ is the dipole–dipole interaction tensor defined by:
\begin{equation}
    T_{ij} =
    \frac{1}{r_{ij}^3} I - \frac{3}{r_{ij}^5}
    \left[ \begin{matrix} x^2 & xy & xz \\ yx & y^2 & yz \\ zx & zy & z^2 \end{matrix} \right]
\end{equation}
where $I$ is the identity matrix and $x$, $y$ and $z$ are Cartesian components along the vector between atoms $i$ and $j$ at distance $r_{ij}$.

When polarizable potentials are employed
in \qmmm calculations, an additional energy term arises in the \qmmm interaction:
$E_\textnormal{\qmmm}^\textnormal{Pol}$, which is added to the total energy:
\begin{equation}
    E_\textnormal{\qmmm}^\textnormal{total} =
    E_\textnormal{\qmmm}^\textnormal{Bonded} + E_\textnormal{\qmmm}^\textnormal{Coulomb} +
    E_\textnormal{\qmmm}^\textnormal{VanderWaals} +
    E_\textnormal{\qmmm}^\textnormal{Pol}
\end{equation}
where $E_\textnormal{\qmmm}^\textnormal{Pol}$ describes the explicit polarization, in the case of dipolar force fields, arising from the dipolar interaction between the induced dipoles and the permanent and induced electric field.
This term is calculated via the following formula:
\begin{equation}
    E_\textnormal{\qmmm}^\textnormal{Pol} =
    - \frac{1}{2} \sum_i \alpha_i E_i^0 E_i 
    \label{eq.polqmmm}
\end{equation}
where  $E_i^0$ represents the electrostatic field on atom $i$ arising from the permanent charges, and higher permanent moments (if present), and $E_i$ represents the sum of $E_i^0$ and the electrostatic field on atom $i$ due to the induced dipoles on other sites.

AMOEBA is an example of a polarizable force field that employs the induced dipole formalism and has been implemented for \qmmm simulations. \cite{Kratz2016,Dziedzic2016,Mao2017,Loco2017,Lichem2019,Dziedzic2019}
AMOEBA uses the following functional form:
\begin{eqnarray*}
E_\textnormal{int}^{\textnormal{AMOEBA}}(\mathbf{r}^N)&=&\sum_{\textnormal{bonds}}k_i^{b}(l_i-l_{i,0})^2[1+2.55(l_i-l_{i,0})+3.793125(l_i-l_{i,0})^2] \\
&+&\sum_{\textnormal{angles}}k_i^{\theta}(\theta_i-\theta_{i,0})^2[1+0.014(\theta_i-\theta_{i,0})+5.6\times10^{-5}(\theta_i-\theta_{i,0})^2\\
&+&7.0\times10^{-7}(\theta_i-\theta_{i,0})^3+2.2\times10^{-8}(\theta_i-\theta_{i,0})^4] \\
&+&\sum_{\textnormal{torsions}}\frac{V_n}{2}(1+ \cos(n\omega -\gamma)) + \sum_{\textnormal{oop}}0.02191414k_{\chi}\chi^2 \\
&+&\sum_{\textnormal{str-bend}}k_{sb}(b_i-b_{i,0})(\theta_i-\theta_{i,0})+\sum_{\textnormal{PI-tor}}U_{\textnormal{PI-tor},i}+\sum_{\textnormal{tor-tor}}U_{\textnormal{tor-tor},i} \\
&+&\frac{1}{2}\left[\sum_{i<j}^NM_i T_{ij} M_j+\sum_{i<j}^N\epsilon_{ij}\left(\frac{1+\delta}{\rho_{ij}+\delta}\right)^{n-m}\left(\frac{1+\gamma}{\rho_{ij}^m+\gamma}-2\right)\right]+\frac{1}{2}\sum_{i}^N\mu_i^{ind} E_{i}
\end{eqnarray*}


Given that the induced dipoles depend on the electric fields, Eq.
\eqref{eq.polqmmm} can be re--written as:
\begin{equation}
     E_\textnormal{QM/MM}^{\textnormal{Pol}} =
     \frac{1}{2} \sum_i 
     \sum_{j \neq i} \mu_i T_{ij} \mu_j
     - \sum_i \mu_i (E_i^\textnormal{MM} + E_i^\textnormal{QM})
     \label{eq.polqmmm2}
\end{equation}
where the first term corresponds to the induced dipole interaction, and the second corresponds to the interaction of the induced dipoles at site $i$ with the electric field at site $I$ arising from the MM ($E_i^\textnormal{MM}$) and QM ($E_i^\textnormal{QM}$) subsystems.

It is important to note that the equations for the induced dipoles and the QM wave function are coupled.
These equations can be solved iteratively using an SCF approach by computing the induced dipoles at each SCF cycle, resulting in a fully self--consistent (fsc) approach \cite{Loco2016}. Alternatively, it is possible to de--couple the solution
of the iterative dipoles by approximating the MM external field interacting with the QM wavefunction by only including the permanent field arising from the MM environment \cite{Kratz2016}, resulting in a partially self--consistent (psc) \qmmm polarization approach.

Force fields that explicitly consider polarization changes in response to the chemical environment offer a more accurate description of electrostatic interactions.
In fact, it is known that polarization effects become more relevant in highly charged systems \cite{Li2015}.
The dipole moment in a molecule is known to change significantly according to the state of aggregation.
A polarizable force field has the versatility to reproduce both gas phase and liquid phase properties simultaneously.
For example, the non-polarizable TIP3P \cite{Jorgensen1983} additive model calculates a hydrogen bond energy for a water dimer greater than the accepted value.
In contrast, polarizable models produce energy closer to the values calculated with ab initio methods \cite{Lopes2009}.
In addition, distributed multipoles introduce charge density anisotropy that is lost in isotropic point charge distributions.
Table \ref{table:amoeba_tip3p} shows the mean absolute error for 10 water dimers calculated with TIP3P and AMOEBA \cite{Kratz2016}.
The results show that the AMOEBA polarizable force field describes hydrogen bonds better than the TIP3P model.

\begin{table}[ht]
\caption{Mean absolute error of the binding energy for 10 water dimers compared with BSSE corrected binding energies from PBE0/aug- cc-pVTZ (PBE0/6-311++G(d,p)). Adapted with permission from E. G. Kratz, A. R. Walker, L. Lagard{\`e}re, F. Lipparini, J.-P. Piquemal, G. Andrés Cisneros J. Comput. Chem. 2016, 37, 1019–1029. Copyright 2016 American Chemical Society}
\begin{tabular}{p{2.5cm}p{2.5cm}p{2.5cm}p{2.5cm}}
\hline
&  & \multicolumn{2}{c}{MAE, $E_{\textnormal{bind}}$} \\ \cline{3-4}
QM & MM & meV & kcal/mol \\
PBE0 & AMOEBA & 19.02 (25.54) & 0.4385 (0.5889) \\
PBE0 & TIP3P & 32.89 (27.75) & 0.7584 (0.6398) \\
- & AMOEBA & 9.26 (36.62) & 0.2136 (0.8446) \\
- & TIP3P & 44.84 (27.19) & 1.0340 (0.6270)
\end{tabular}%
\label{table:amoeba_tip3p}
\end{table}



\begin{figure}[H]

\begin{center}

\includegraphics[width=0.8\textwidth]{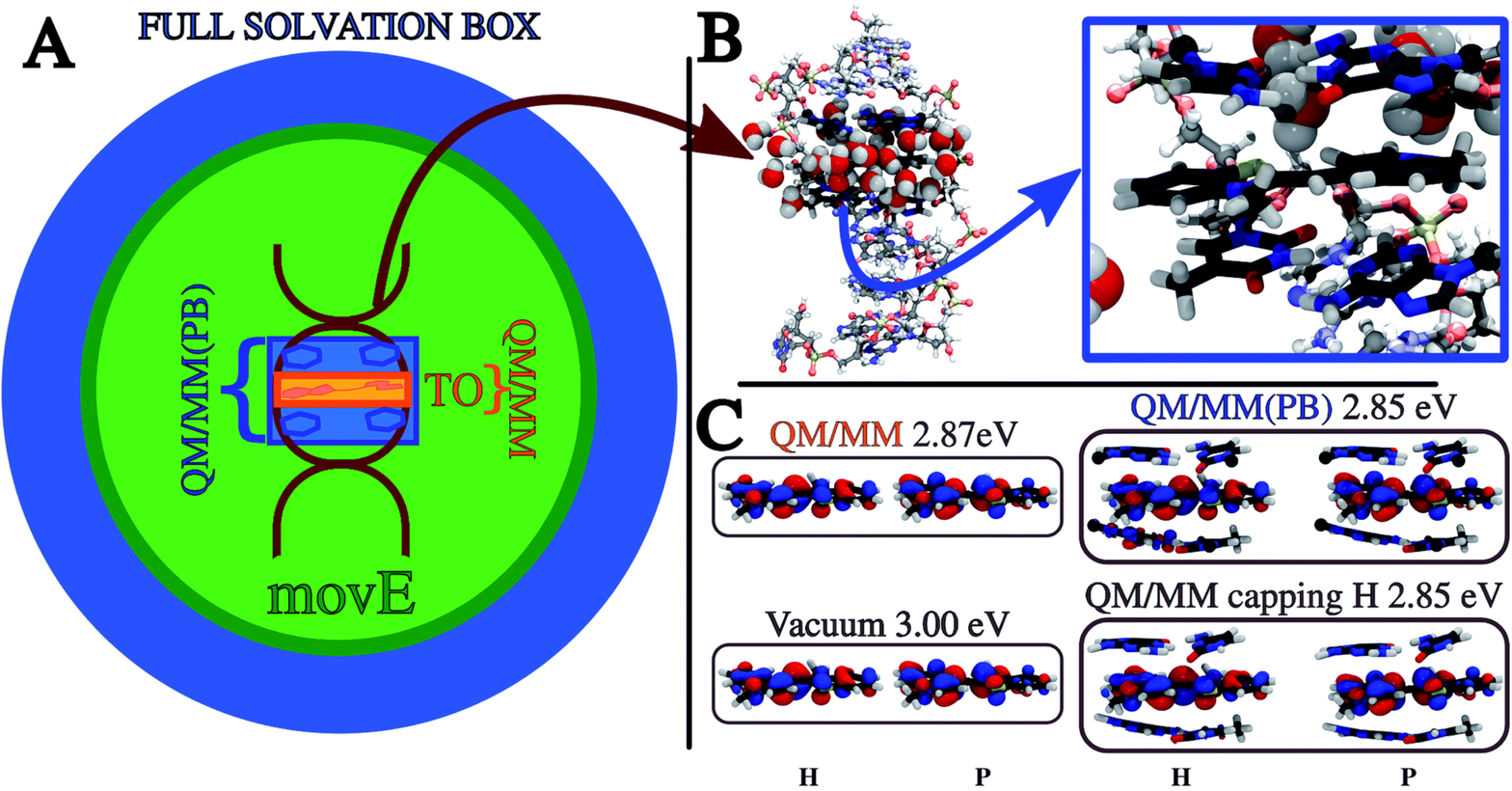}

\caption{(A) Pictorial representation of the TO dye buried in the DNA double helix, embedded in a sphere of water (16500 atoms). The different colors represent the differences for each of the QM/MM partition schemes employed
(B) The DNA structure is highlighted in a ball-and-stick representation, leaving the first water solvation shell around the TO dye. The QM subsystem defined in this work, tagged as QM/MM(PB) and made up of the TO dye and the four closest nucleobases (NBs) is zoomed out and represented in licorice style in the blue square. 
(C) NTOs and excitation energies relative to the $\pi\pi^*$ bright excitation of the TO dye embedded in different environments, from vacuum to the QM/MM(PB) partition scheme. 
H and P denote the hole and particle, respectively.
Reproduction from D. Loco, L. Lagardère, G. A. Cisneros, G. Scalmani, M. Frisch, F. Lipparini, B. Mennucci and J.P. Piquemal, Chem. Sci., 2019, 10, 7200 -  Published by The Royal Society of Chemistry. \cite{Loco2019}} 
\label{fig:thiazole-orange}
\end{center}
\end{figure}

Another recent example of the importance of polarizable
environments in \qmmm is shown in the work of Loco \etal \cite{Loco2019}.
In this work the \qm/\amoeba method, which interfaces the \gaussian \cite{Gaussian} and \tinker/\tinkerhp \cite{TINKER-HP} packages has been used to compute excitations properties on the Thiazole Orange dye (TO) intercalated in DNA solvated in water (\figrff{fig:thiazole-orange}).
Several different types of \qmmm molecular dynamics (MD) simulations have been conducted, here we will focus on two of them. First, the \qm subsystem is only composed of the TO dye (denoted 
\qmmm) as previously done by Loco \etal \cite{Loco2018}. Second, the \qm subsystem is composed of the TO dye and two pairs of nucleobases (denoted \qmmm(PB)). 
The calculated results indicate charge-transfer intruder states for the larger \qm subsystem (\qmmm(PB)), leading to the delocalization of the TO $\rm \pi\pi^{*}$ transition on the \qm nucleobases. Though, we observe a small difference of the excitation energy value between \qmmm and \qmmm(PB) when using the AMOEBA polarizable force field. However, complementary studies, using the non-polarizable AMBER99sb force field \cite{Hornak2006} and the TIP3P water model, have shown a non-negligible difference of the excitation energy between \qmmm and \qmmm(PB). Therefore, computed excitation energies from polarizable \qmmm are less sensitive to the choice of the \qm subsystem than classical \qmmm simulations. Embedding polarization effects are a key factor not only for \qmmm calculations but also to compute gas and condensed phase properties. 

\section{\qmmm Boundary Considerations}
\label{sec:qmmm-boundary}

The combination of quantum and classical methods results in short--
and long--range boundaries that require special consideration. 
In the short--range regime, the junction between the QM subsystem and the MM 
environment requires care in cases where covalent bonds are present
across the \qmmm boundary. There are scenarios where, even
if no chemical bonds are cut, it is necessary to pay special attention
to this boundary, for example if molecular exchange between the QM
and MM regions is possible or likely. For long--range interactions, it
is necessary to consider the role of these effects on the \qmmm energies
and forces. In this section we provide a brief review of approaches
that have been developed to address both types of scenarios.

\subsection{Boundary Between QM and MM subsystems}
\label{closerange}

One of the challenges in \qmmm is the description of atoms at the boundary between \qm and \mm subsystems involving covalent bonds. 
In general, atoms that are important in the reactive process are described at the \qm level of theory, while the remaining atoms are described at the \mm level of theory. In the case where only non-covalent interactions are involved, the separation between the \qm and \mm subsystems is straightforward. In the cases where a covalent bond (or more) need to
be cut to create the two subsystems, the atoms involved in the
covalent bond, and atoms near the boundary, need then to be treated with care.
Indeed, the two different levels of theories used in \qmmm must be coupled to describe atoms and accurately compute the energy of the system. The challenging description of \qmmm boundary atoms have been addressed in different manners leading to the development of various approaches, some
of which are mentioned below.


\begin{figure}[t]

\begin{center}

\subfloat[]{\includegraphics[scale=0.2]{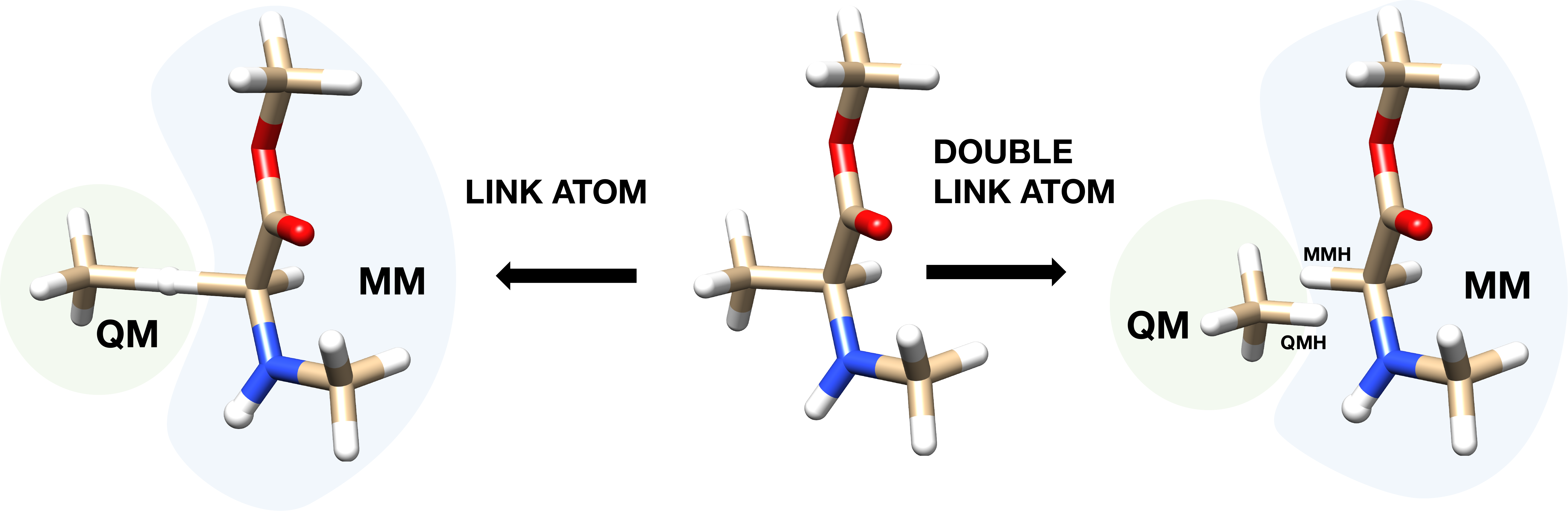}
\label{fig-link-atom-et-double}}

\subfloat[]{\includegraphics[scale=0.25]{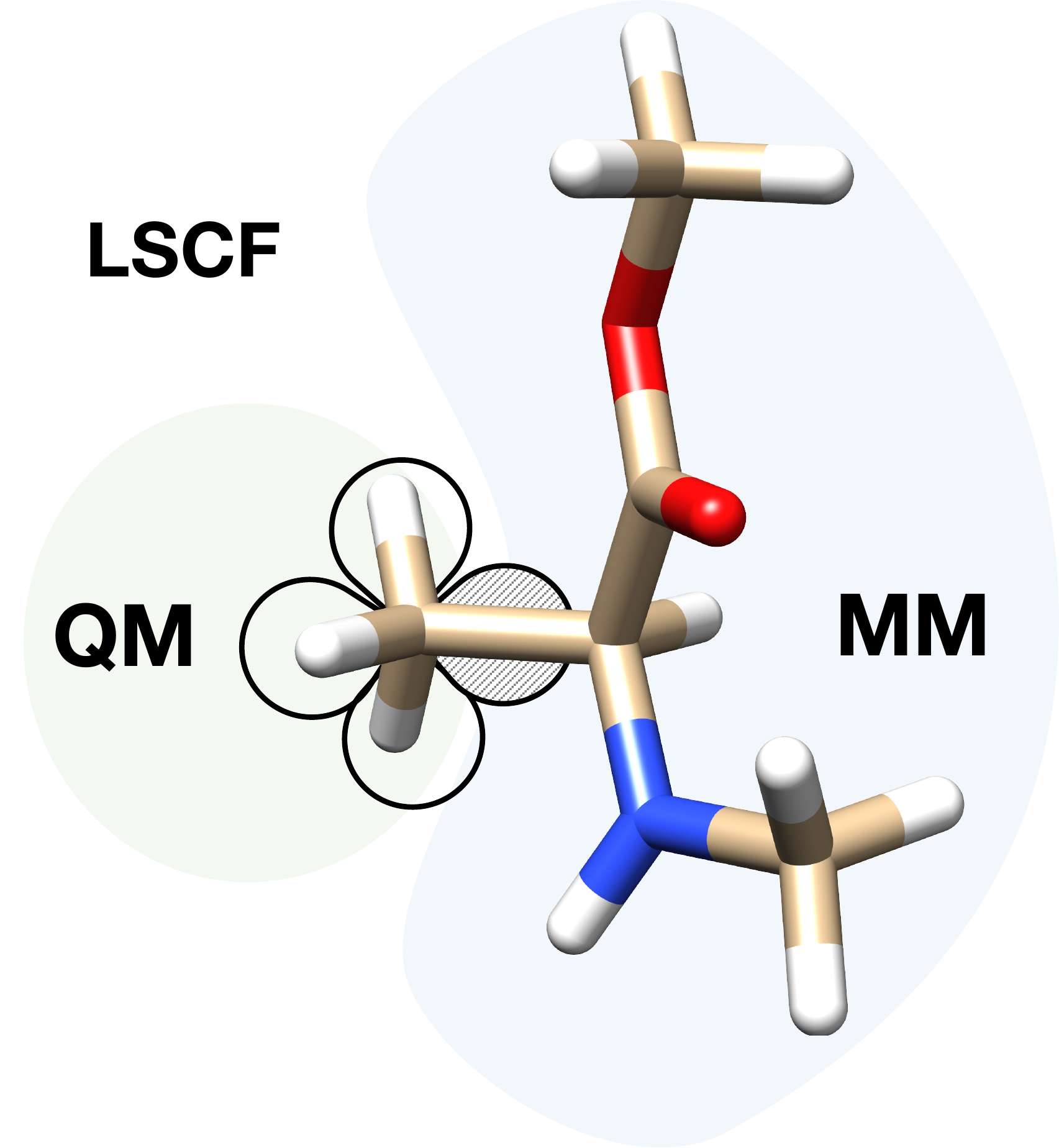}
\label{fig-lscf}}
\hspace*{1cm}
\subfloat[]{\includegraphics[scale=0.25]{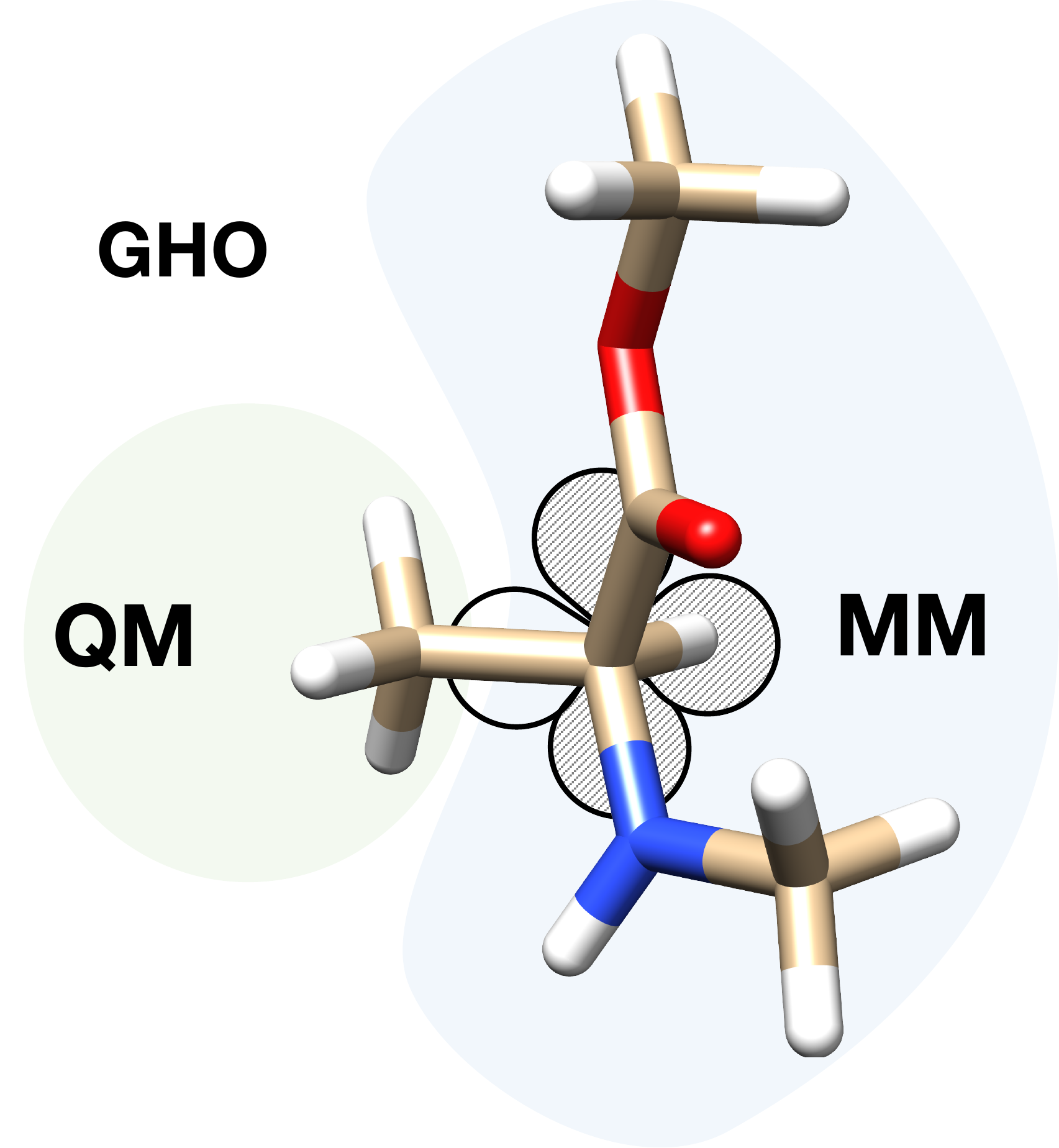}
\label{fig-gho}}

\caption{Representation of  
\protect\subref{fig-link-atom-et-double} Link Atom approaches: simple link atom \cite{Field1990,Singh1986} (left) and Double Link Atom (right) \cite{Das2002}, and frozen localized orbitals where gray orbitals are kept frozen while plain white orbitals are optimized during \qm optimization : 
\protect\subref{fig-lscf} LSCF \cite{Monard1996,Ferenczy1992,Thery1994,Assfeld1996,Monari2013,Ferre2002}, \protect\subref{fig-gho} GHO methods \cite{JialiGao1998,Amara2000}.}

\label{fig:link-frz}
\end{center}
\end{figure}



The link atom is the most commonly used method due to its simplicity and
ease of implementation \cite{Field1990,Singh1986}. The idea is to add a supplementary atom (i.e link atom), generally an hydrogen atom, to fill
the empty valence in the \qm subsystem arising for each cut covalent bond (\figrff{fig-link-atom-et-double}). The link atom is then only described at the \qm level, resulting in the addition of three extra degrees of freedom. This can result in modifications of the electronic structure and the chemical properties of the \qm subsystem. Moreover, the closeness between the link atom and the charge of the \mm atoms may induce an over-polarization from the \mm towards the \qm subsystem. In order to overcome those issues, improvements
have been suggested, such as the double link atom (DLA) \cite{Das2002} among others. In the DLA method, two supplementary atoms, usually hydrogen atoms, are linked to both atoms of the broken bond to create the \qm and \mm subsystems. The idea here is to overlap both \qm and \mm link atoms (QMH and MMH) to ensure an overall neutral net charge, avoiding then an over-polarization from the \mm towards the \qm (\figrff{fig-link-atom-et-double}). Also, it is possible to use the double link-atom method with a point charge or a single s--type Gaussian function
on the boundary atoms.  In the latter, the charges are delocalized, providing a better description of the  charge density distribution of the \mm boundary atoms. 

Methods using frozen localized, hybrid orbitals have been developed to describe \qmmm boundary atoms, following pioneering work from Warshel and Levitt \cite{Warshel1976}. Later, Rivail and coworkers \cite{Monard1996,Ferenczy1992,Thery1994,Assfeld1996,Monari2013,Ferre2002} have developed the Localized Self-Consistent Field (LSCF) method. The idea is to describe the \qm atom of the broken bond using a set of hybrid orbitals. The orbital pointing towards the \mm atom (shaded gray orbital in \figrff{fig-lscf}) is kept frozen during the \qm SCF optimization, while other orbitals (plain white orbitals in \figrff{fig-lscf}) are optimized with all \qm orbitals. In a related approach, the Generalized Hybrid Orbital (GHO) method developed by Gao \etal \cite{JialiGao1998,Amara2000}, the set of orbitals is placed on the \mm atom of the broken bond. In this case, the hybrid orbital pointing towards the \qm atom is optimized with all the \qm orbitals (plain white orbital in \figrff{fig-gho}) while the other orbitals on the \mm atom are kept frozen during the SCF optimization procedure. Overall, the use of frozen orbitals does not alter the electronic and chemical properties of the system. However, it requires additional parametrization steps of the orbitals depending on the nature of the bond to be cut.

\begin{figure}[H]

\begin{center}
\subfloat[]{\includegraphics[scale=0.25]{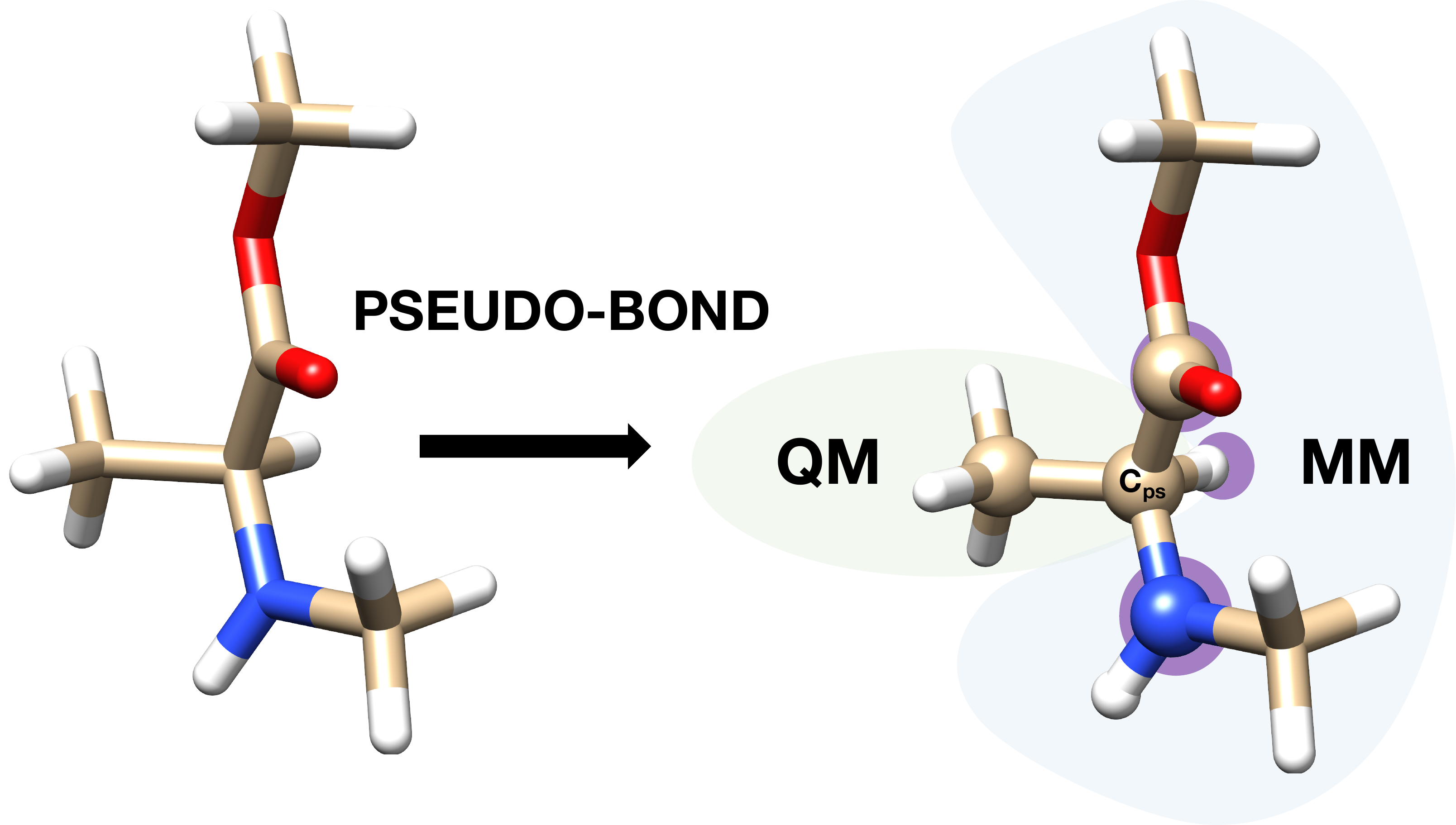} 
\label{fig-pb}}

\subfloat[]{\includegraphics[scale=0.25]{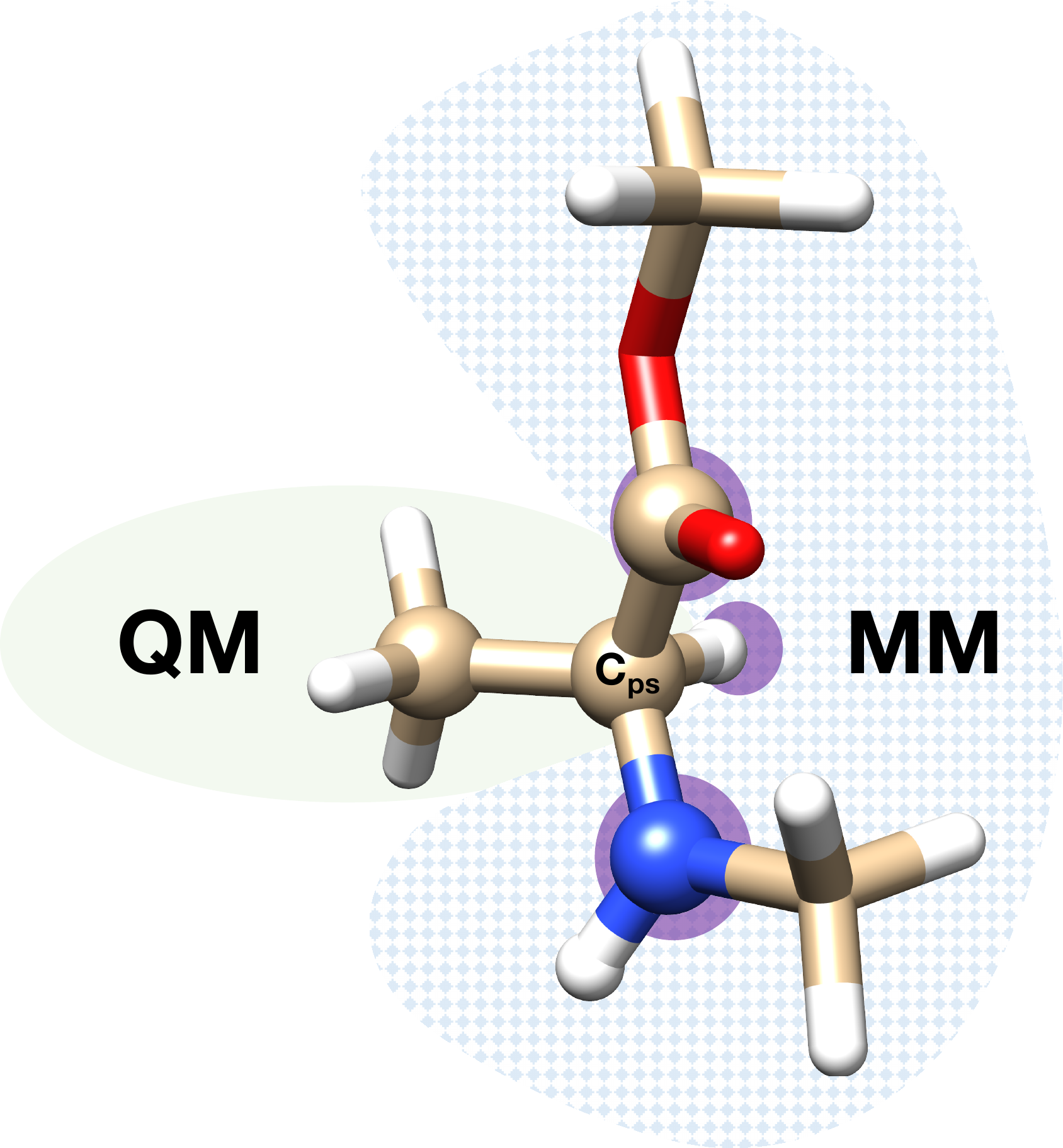}
\label{fig-pb-lichem-A}}
\hspace*{1cm}
\subfloat[]{\includegraphics[scale=0.25]{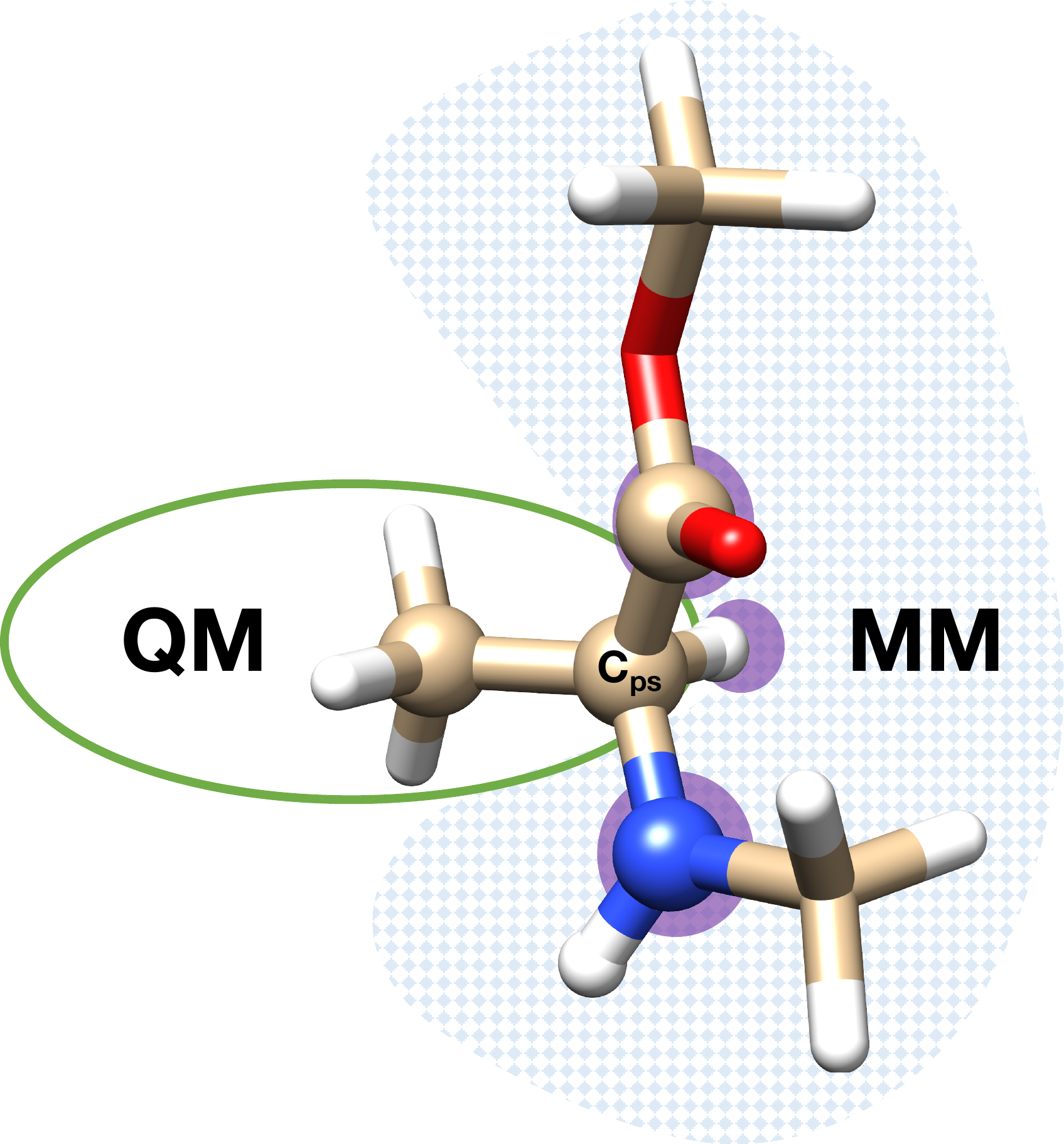}
\label{fig-pb-lichem-B}}

\subfloat[]{\includegraphics[scale=0.25]{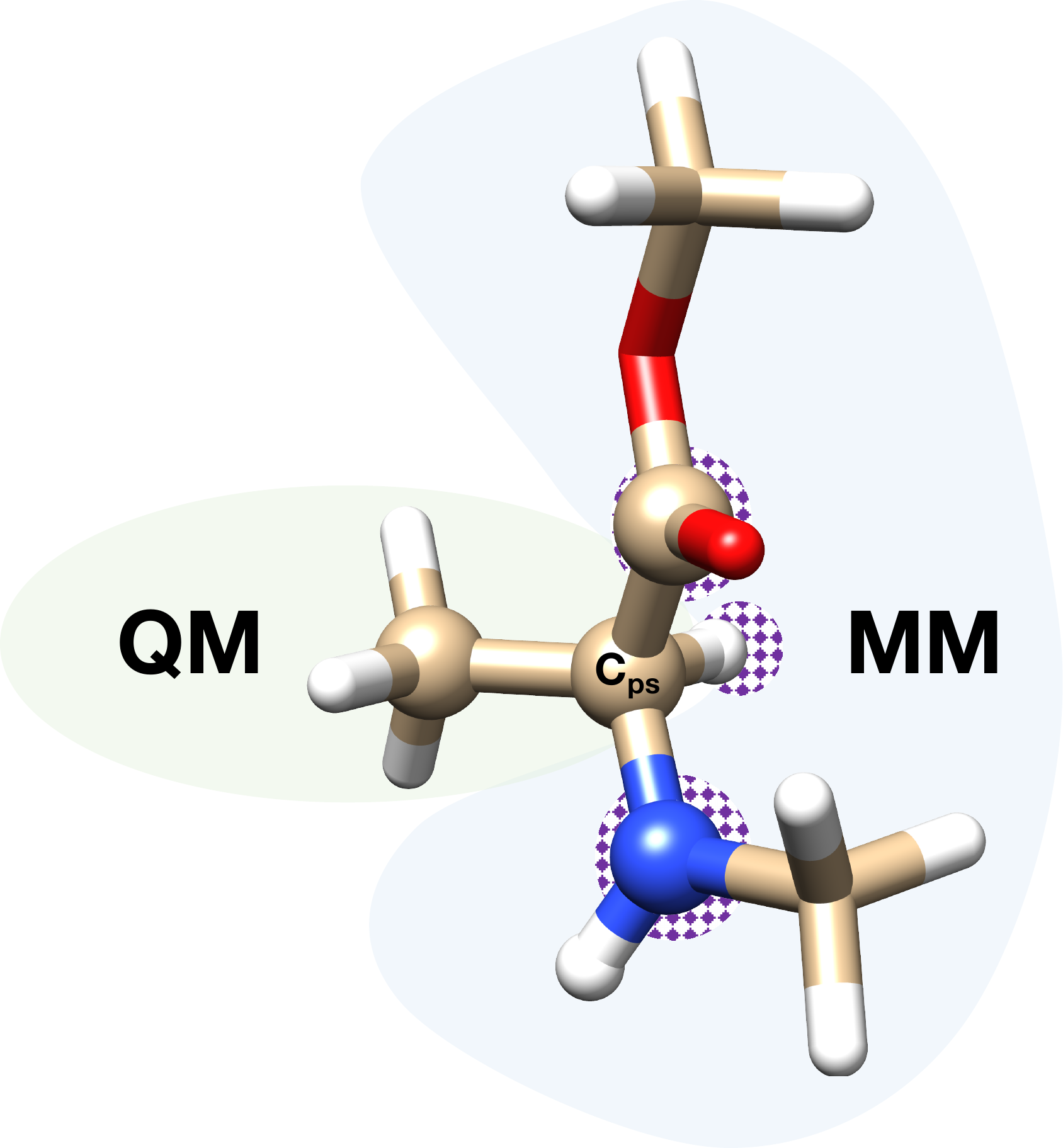}
\label{fig-pb-lichem-C}}
\hspace*{1cm}
\subfloat[]{\includegraphics[scale=0.25]{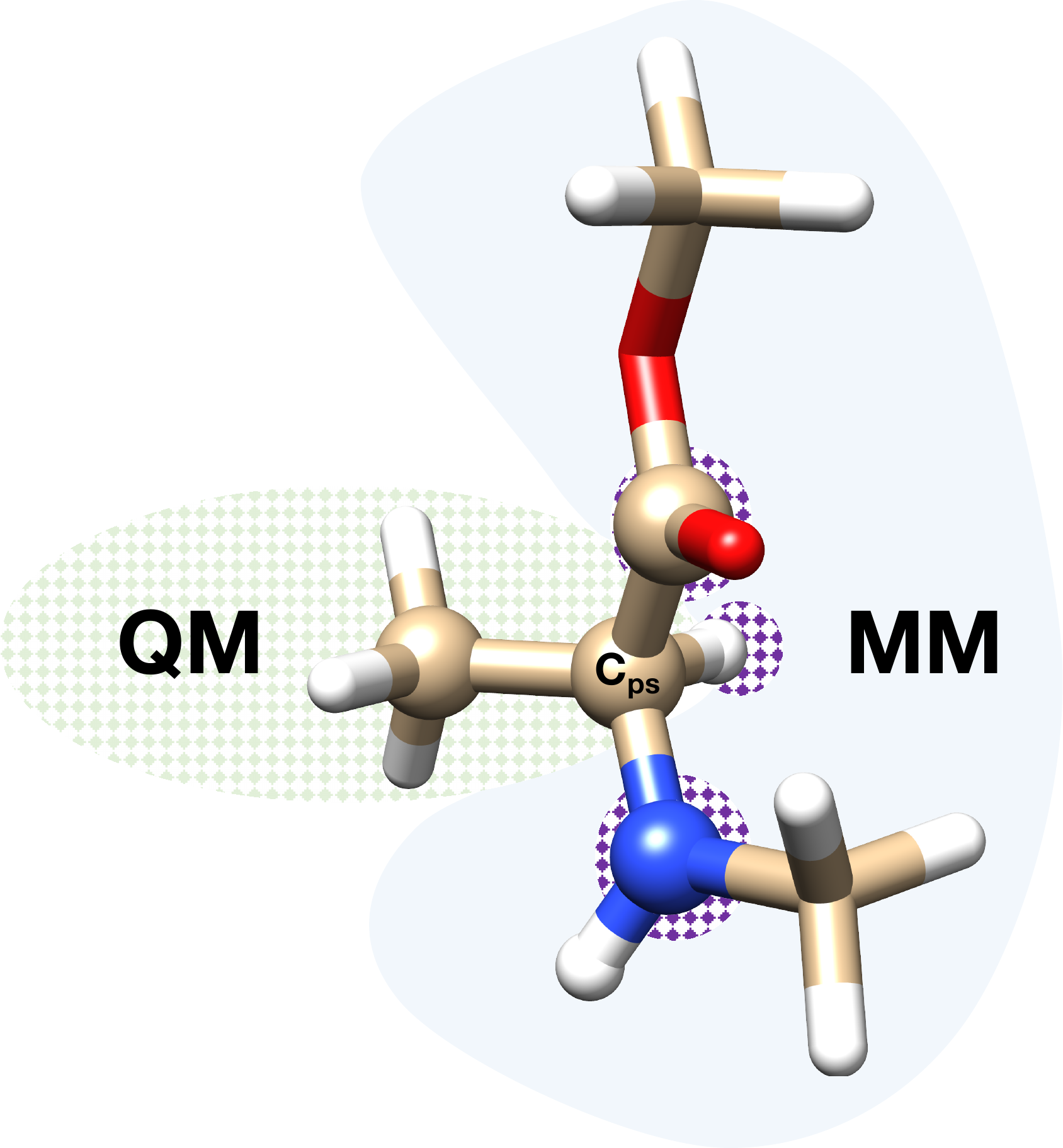}
\label{fig-pb-lichem-D}}

\caption{\protect\subref{fig-pb} Molecular schematic representation
of the pseudobond approach for polarizable \qmmm optimization, with QM (MM)
atoms in the green(blue) shaded regions, and boundary atoms shaded in
purple. In this approach, a polarizable \qmmm involves four calculations:
\protect\subref{fig-pb-lichem-A} QM polarized by the static
MM field (blue shaded region), \protect\subref{fig-pb-lichem-B} MM without
charges on QM (green circle) or boundary atoms (purple shaded atoms), \protect\subref{fig-pb-lichem-C} MM polarization including
static (approximate) QM field (green and purple shaded regions), \protect\subref{fig-pb-lichem-D} MM optimization with static field from QM and boundary atoms (green and purple shaded regions) \cite{Kratz2016}.}

\label{fig:pb-lichem}
\end{center}
\end{figure}

The pseudo-bond approach involves the use of "taylor--made" pseudopotentials. In this method the \qm atom of the broken bond is described using effective core potentials (ECP) specifically parametrized
to reproduce the environment of the covalent bond. The \qm atom is denoted as C$_{ps}$ and is located in the \qm region (\figrff{fig-pb}). Pseudo-bonds should mimic the original chemical and structural properties of the broken bond between the \qm and \mm subsystems. Following Zhang \etal previous works \cite{Zhang1999, Zhang2005}, Parks \etal \cite{Parks2008} have developed new parameters for pseudo-bonds for common bonds found in biological systems 
such as: 
C$_{ps}$($sp^3$)-C($sp^3$), C$_{ps}$($sp^3$)-C($sp^2$,carbonyl), and C$_{ps}$($sp^3$)-N($sp^3$). The parametrization of the ECPs are independent of the force fields and the latest pseudobonds also include a minimal basis-set, which overcomes previous dependence of the pseudobond to specific basis sets. In parallel, Dilabio \etal \cite{DiLabio2002} have developed quantum capping potentials, which include additional terms compare to ECP as spherical screening and Pauli repulsion, which contribute then to the refinement of pseudo-bonds.

In the case of polarizable \qmmm, special care needs to be taken to address
the polarization interaction across the \qmmm boundary. The
pseudobond approach has been extended to encompass polarization effects
\cite{Kratz2016}. In addition to using pseudobonds, boundary atoms (purple 
shaded surface in \figrff{fig:pb-lichem}) need to be considered
to compute a \qmmm optimization, which is performed by means of four energy 
calculations such as : \qm energy in the \mm static field 
(\figrff{fig-pb-lichem-A}), \mm energy with no charge on the \qm atoms and 
boundary atoms (\figrff{fig-pb-lichem-B}), \mm polarization energy with 
charges on the \qm atoms (\figrff{fig-pb-lichem-C}), \mm optimization with no 
polarization from the \qm subsystem, pseudo-bond and boundary atoms 
(\figrff{fig-pb-lichem-D}). This allow then to avoid the over-polarization from the \mm towards the \qm, 
Kratz \etal \cite{Kratz2016} have demonstrated the accuracy of using this approach on oligopeptide chains in the gas phase with \lichem \cite{Kratz2016}. 

\newpage


In another recent example, Das \etal \cite{DasR2019} studied the polarization effect of uracil DNA glycosylase (UDG) on substrates leading to a ''tautomeric strain'' phenomena. UDG belongs to the excision repair enzyme family which its role is to cleave the N$_1$-C$_{1}'$ to remove mutagenic uracil from DNA. The activation of the N$_1$-C$_{1}'$ bond has been investigated actively both the focus of experimental and computational 
studies because uracil is not considered as the best of leaving group, 
giving rise to questions about the reaction mechanism. 

\begin{figure}[H]

\captionsetup{justification=raggedright,singlelinecheck=false}
\begin{center}

\includegraphics[width=0.7\textwidth]{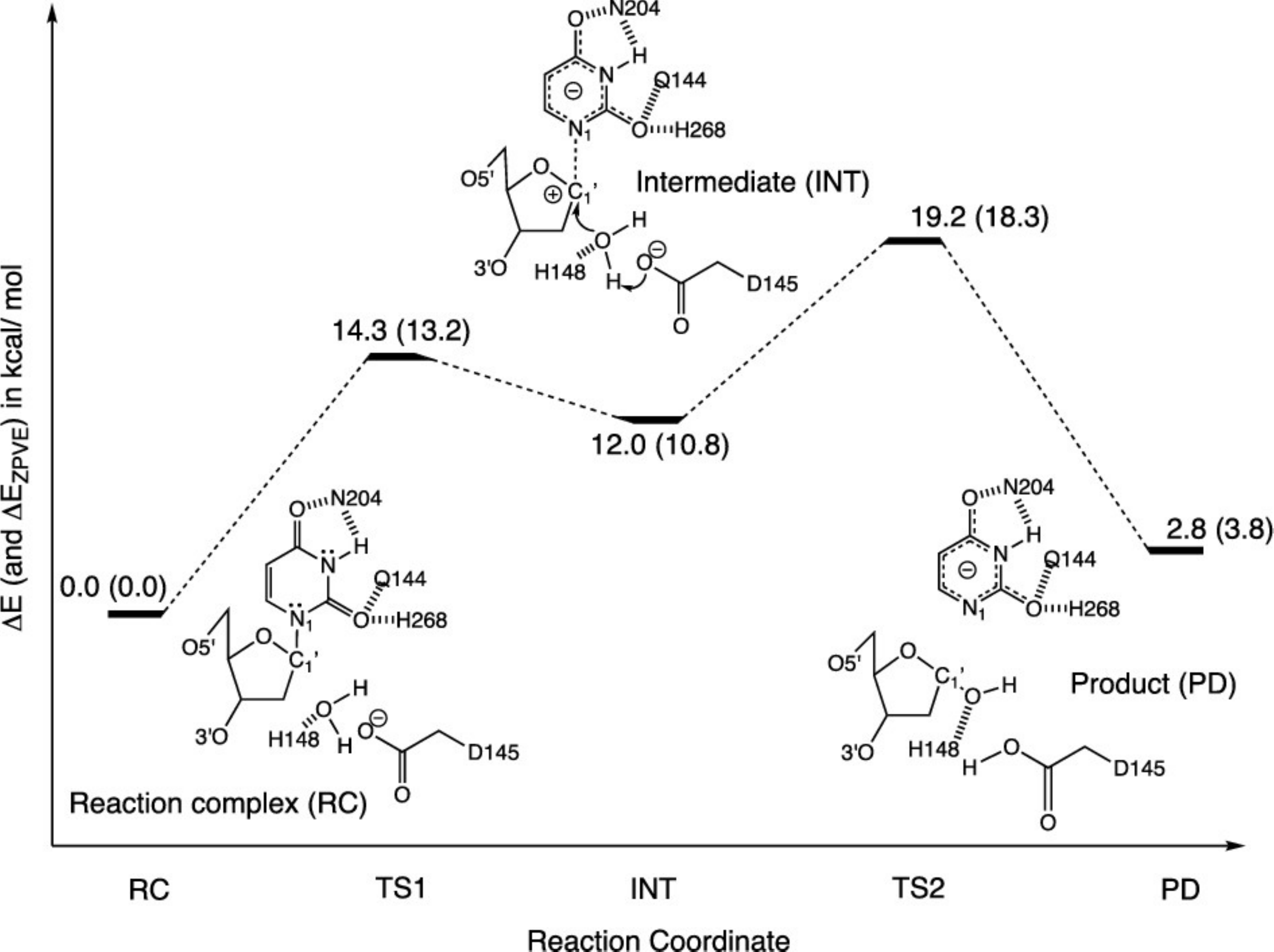}

\caption{Relative energies ($\Delta$E, in kcal/mol) of computed stationary points along the stepwise dissociative pathway of UDG, based on a constrained and truncated QM model of the UDG active site. Values corrected for zero-point energy vibration (ZPVE) are shown in parentheses ($\Delta E_{ZPVE}$, in kcal/mol). Geometries of all stationary points were optimized at the $\omega$B97X-D/6-31+G(d) level.
Reprinted with permission from J. Am. Chem. Soc. 2019, 141, 35, 13739–13743. Copyright 2019 American Chemical Society. \cite{DasR2019}} 
\label{fig:udg-path}
\end{center}
\end{figure}

Das \etal performed QM calculations on a cluster model based on the
active site as well as \qmmm calculations on the full UDG system 
including eight pseudobonds to investigate the effect of the environment
on the aromaticity of uracil, and how this affects the reactivity  \cite{DasR2019}. Based on the results from both approaches, a step--wise dissociative mechanism for UDG was proposed (Figure
\ref{fig:udg-path}).  Interestingly, the active site of UDG includes several nearby residues interacting with UDG by means of hydrogen bonds. Calculated aromaticity indicators (NICS and HOMED)
for uracil in the cluster model suggest that the uracil experiences an
increase in aromaticity compared with uracil in the gas phase, to values
similar to higher energy uracil tautomers. Including the full
polarized protein environment increases the aromaticity of uracil even more.
This ''tautomeric strain'' enables the reduction of the activation barrier to cleave uracil from UDG thanks to the polarization effects \cite{DasR2019}.

\subsection{Long-Range Electrostatics in \qmmm}
\label{sec:qmmm-elec}
Long-range interactions in biomolecular and condensed--phase simulations play an important role that needs to be accounted for in order to properly represent a bulk system.
One approach to simulate a bulk system is by the use of periodic boundary conditions (PBC). 
A popular approach to include long--range effects under PBC in classical simulations is the smooth particle mesh Ewald (sPME) approach \cite{Essmann1995}.
The sPME approach relies on an approximation of the Ewald summation method \cite{Ewald1921} that uses the fast Fourier transform approach, resulting in $\mathcal{O}(N \log N)$ scaling.




Briefly, for a system with $N$ point charges, the electrostatic energy of a system under PBC is given by:
\begin{equation}
    E_\textnormal{elst} =
    \frac{1}{2} \sum_n^* \sum_{i,j=1}^N \frac{q_i q_j}{|\textbf{r}_i -\textbf{r}_j+\textbf{n}|^2}
    \label{electrostatic_PBC}
\end{equation}
where the asterisk over the sum denotes that the terms with $\textbf{n}=0$ and either $j=i$ are omitted.


The direct calculation of the electrostatic energy in Eq. \eqref{electrostatic_PBC} is $\mathcal{O}(N^2)$ and additionally, is only conditionally convergent.
The Ewald method \cite{Ewald1921} divides the electrostatic interaction into a sum of three contributions: direct, reciprocal, and correction terms,
\begin{equation}
    E_\textnormal{elst} = E_\textnormal{direct} + E_\textnormal{recip} + E_\textnormal{corr}
    \label{ewald_summation}
\end{equation}
The direct term is given by
\begin{equation}
    E_\textnormal{dir} =
    \frac{1}{2} \sum_n^* \sum_{i,j=1}^N q_i q_j \left( \frac{\textnormal{erfc} (\beta|\textbf{r}_{j}-\textbf{r}_j+\textbf{n}|)}{|\textbf{r}_{j}-\textbf{r}_i+\textbf{n}|}\right)
\end{equation}
where $\beta$ is the Ewald splitting parameter. The reciprocal term is defined as
\begin{equation}
    E_\textnormal{recip} =
    \frac{1}{2 \pi V} \sum_{m \neq 0} \frac{\textnormal{exp}(-\pi^2 \textbf{m}^2 / \beta^2)}{\textbf{m}^2} S(\textbf{m}) S(\textbf{-m})
\end{equation}
where $V=\textbf{a}_1\cdot\textbf{a}_2\times\textbf{a}_3$  is the volume of the box, $\textbf{m}=m_1\textbf{a}_1^*+m_2\textbf{a}_2^*+m_3\textnormal{a}_3^*$ is the lattice in reciprocal space and $S(\textbf{m})$ is the structure factor defined by
\begin{equation}
    S(\textbf{m}) =
    \sum_{j=1}^N q_j \textnormal{exp} (2\pi i m \cdot r_j)
\end{equation}
The correction term in Eq. \eqref{ewald_summation} is composed by the self energy ($E_\textnormal{self})$. 
\begin{equation}
    E_\textnormal{self} =
    -\frac{\beta}{\sqrt{\pi}}  \sum_{i=1}^N q_i^2
    \label{energy_self}
\end{equation}

The above expressions for the correction can be modified to take into account other corrections such as non--neutral unit cells, polarization response, surface terms, etc \cite{Darden2010}.



The Ewald method avoids the conditional convergence in Eq. \eqref{electrostatic_PBC}, albeit the algorithm is $\mathcal{O}(N^2)$.
The sPME and related approaches approximate the reciprocal sum by using a discrete convolution on an
interpolating grid, resulting in a $\mathcal{O}(NlogN)$ algorithm. 
This approach is extremely efficient when a discrete representation, i.e. point charges or point multipoles, is employed, because the discrete charge distribution can be interpolated on a grid in a relatively straightforward fashion \cite{Darden2010}.

The interpolation on a regular grid (or set of grids) of the continuous charge density, such as that from the QM subsystem in \qmmm calculations, presents a challenge.
One way to overcome this challenge is to approximate the continuous QM charge density by a discrete representation (e.g. point charges) as was originally proposed by Nam \etal \cite{Nam2005} for QM/MM--Ewald, and by Walker \etal \cite{Walker2008}.
Giese and York developed the ambient potential composite Ewald approach, which avoids the use of dense Fourier grids for the QM electronic density, and performs a direct interaction of the QM charge density with the reciprocal potential \cite{Giese2016}. The use of an accurate approach to interact the continuous charge density of the QM subsystem with the charges of the MM environment was shown to avoid SCF instabilities and artifacts in PMF calculations. \cite{Giese2016}


Long--range effects can also be approximated by boundary potential models.
In the case of \qmmm calculations several methods have been proposed, including the empirical valence bond (EVB/MM) framework \cite{Warshel1980}, the spherical solvent boundary potential (SSBP) \cite{Beglov1994},
the generalized solvent boundary potential (GSBP) \cite{Im2001}, the solvated macromolecule boundary potential (SMBP) \cite{Benighaus2011}.
One more approach to approximate long--range effects is to use a cutoff method coupled with switching or shifting functions, without taking the long--range effects explicitly into account.
\cite{McCann2013,Acevedo2014,Acevedo2014b}

Another method that has been developed for both point--charge, and multipolar/polarizable \qmmm calculations is the QM/MM Long-Range Electrostatic Correction (QM/MM--LREC). 
The QM/MM--LREC scheme uses a minimum image convention and a switching and smoothing function to avoid edge effects and incorporate long--range electrostatic interactions, while at the same time matching the energies and forces with sPME levels of accuracy \cite{Fang2015,Kratz2016}.

The minimum image convention method considers the interactions between each particle $i$ with all particles $j$ within a cutoff radius $R_{\textnormal{cut}}$.
If the distance $r_{ij}$ between the particles $i$ and $j$ is less than half the length of the primary unit cell then the same image is considered, on the other hand, if the distance between the particles $i$ and $j$ is larger, the neighboring image is used.

In the QM/MM-LREC scheme, each particle $i$ interacts with every particle $j$ within a certain cutoff distance $r_{ij}$.
This leads to a modified Coulomb interaction such that
\begin{equation}
     E(q_i,q_j,r_{ij}) = \frac{q_i q_j}{r_{ij}} f(r_{ij}',s)
\end{equation}
where
\begin{equation}
    f(r_{ij}',s)=
    \left[1 - \left( 2r_{ij}'^3 - 3r_{ij}'^2 +1 \right)^s \right]
\end{equation}
and
\begin{equation}
    r_{ij}'= \left( \frac{R_{\textnormal{cut}}-r_{ij}}{R_{\textnormal{cut}}} \right)
\end{equation}

\begin{figure}[h]
    \centering
    \includegraphics[width=0.7\textwidth]{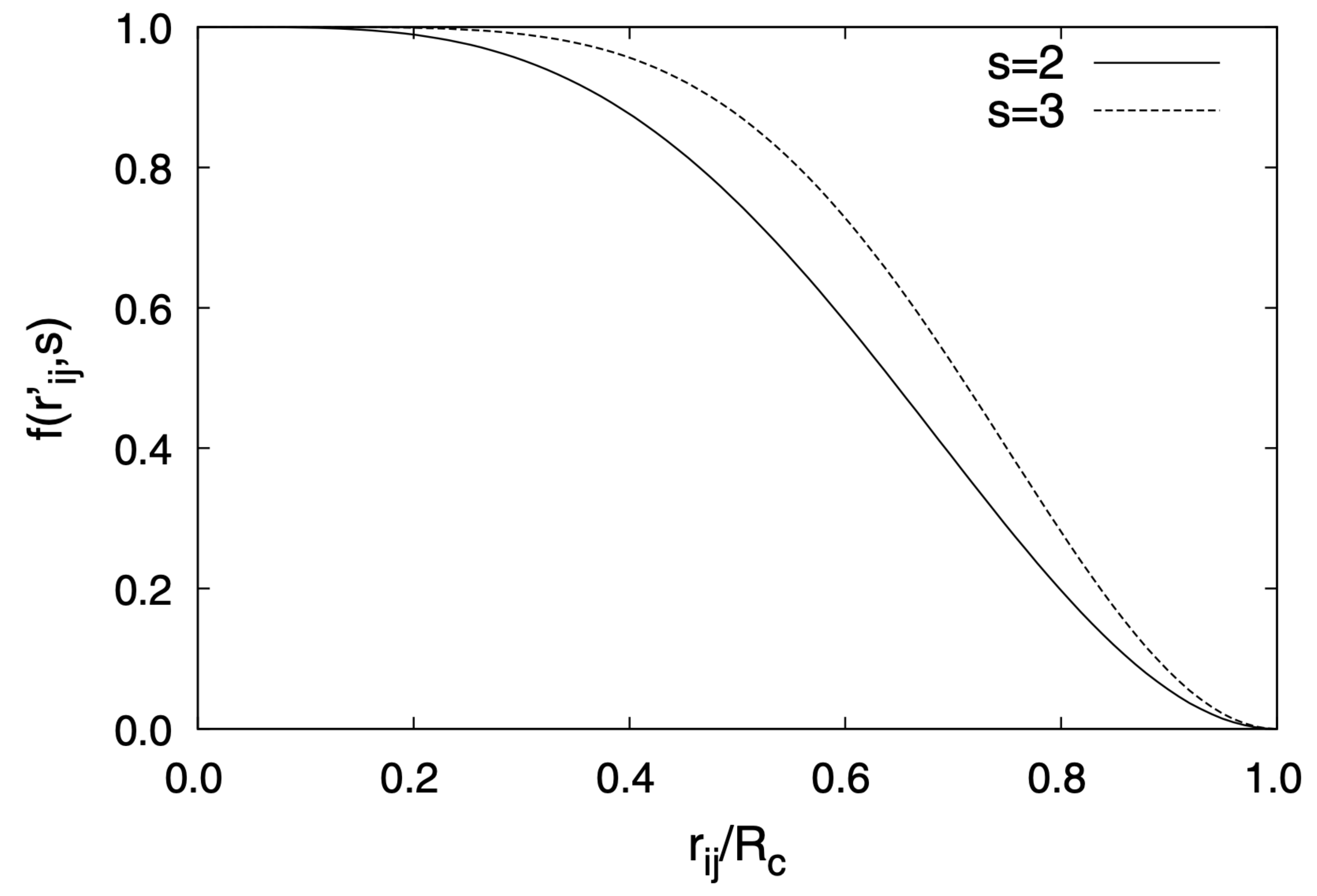}
    \caption{LREC damping function with s = 2 and s = 3. Increasing the exponent moves the inflection point of the LREC function closer to the cutoff radius. Reprinted with permision from Kratz, E.G., Duke, R.E. and Cisneros, G.A. Long-range electrostatic corrections in multipolar/polarizable QM/MM simulations. Theor Chem Acc 135, 166 (2016).}
    \label{fig:lrec}
\end{figure}

Here $s$ is an adjustable integer exponent controlling the decay of the damping function (see \figrff{fig:lrec}).
The LREC scheme avoids the need to approximate the QM charge density for reciprocal space by considering only
the first replicas around the primary cell.
It also has the advantage that analytical Hessians can be calculated analytically, and the switching function $f(r'_{ij},s)$ can be employed for any discrete embedding field including monopoles and higher order multipoles.
This approach has been recently extended to a hybrid approach allowing a reduction in the cutoff distance \cite{Pan2018}.

\section{QM/MM with Advanced Potentials}
\label{sec:qmmm-adv-pot}
The QM/MM approaches described so far only involve electrostatic 
and polarization embedding of the QM wavefunction \eqref{eq.polqmmm}.
The inclusion of explicit polarization improves the description of 
the MM environment, however, other effects are approximated by
calculating them via the classical terms, e.g. Van der Waals, \cite{Riccardi2004}
or are neglected altogether, e.g. charge transfer.

Several approaches have been developed to include a more 
accurate  description of the MM 
environment. One recently proposed approach is to employ potentials
based on the many body decomposition. One such
family of potentials is the MB-pol and MB-DFT
potentials \cite{Reddy1945042016}.
We have recently described a new QM/MM implementation
involving MB-pol and MB-DFT within LICHEM
and within Gaussian to compare partially
self--consistent (psc) and fully self
consistent (fsc) polarizable QM/MM approaches
\cite{Lambros2020MBPolqmmm}.
These implementations include explicit Coulomb and
polarization embedding of the QM wavefunction
and a many-body potential representation
of the MM environment. The use of MB-pol
and MB-DFT allows for a very accurate
representation of the environment. This results
in various advantages such as no longer needing
additional functions or approximations to ensure continuous energy and forces in adaptive \qmmm calculations \cite{Duster2018,Watanabe2019}.

Another approach to improve the 
description of the environment in 
QM/MM simulations is to use advanced
potentials that explicitly include
one or more of the 
missing components in the embedding environment. One approach that allows an 
explicit term--by--term embedding is
by employing potentials that 
include separate terms for each 
physical interaction, such as
the effective fragment potential
(EFP) which can include exchange
and charge transfer \cite{ViquezRojas15492020},
the fragment exchange
potential \cite{Chen40082020},
the QMFF \cite{Giese3830022017}, Exchange fragment
potential \cite{Chen2020},
among others.
See the recent review of QM:MM and QM:QM
embedding approaches by
Jones {\em et al.} for a thorough review \cite{Jones32812020}.

We have developed a QM/MM 
implementation within LICHEM that interfaces PSI4 with GEM where
the MM subsystem is represented by GEM and thus all
MM and QM/MM interactions are calculated using the GEM interaction terms. 
The Gaussian Electrostatic Model (GEM) is a polarizable potential that uses Gaussian auxiliary basis sets (ABSs) to build the molecular electron density \cite{Cisneros2005}. The philosophy for GEM is to
have a density--based force field with errors below 0.2
kcal/mol for  total energy and forces compared with high level
QM reference. This is achieved by using separate contributions
for each of the terms obtained from QM energy decomposition
analysis (EDA): Coulomb, Exchange, polarization, charge transfer
and dispersion. The GEM densities are obtained by fitting QM 
relaxed densities to a set of auxiliary Hermite Gaussian basis sets. 

Two procedures can be used to perform the adjustment of molecular electron densities through Hermite Gaussians: the conventional analytical variational Coulomb adjustment or the numerical adjustment of the molecular electrostatic potential.
The analytical density fitting method relies on the use of Gaussian auxiliary bas functions to expand molecular electron density
\begin{equation}
    \Tilde{\rho} (r) = \sum_k c_k K (r)
\end{equation}
where $\Tilde{\rho}$ is the approximate density and $K(r)$ are Hermite Gaussians.
The expansion coefficients $c_k$ can be obtained by minimizing the self-energy of the error in the density according to some metric
\begin{equation}
    E_\textnormal{self} =
    \langle \rho(r)-\Tilde{\rho}(r)|\hat{O}|\rho(r)-\Tilde{\rho}(r)\rangle
    \label{metric}
\end{equation}
Different operators $\hat{O}$ can be used, including the coulomb operator $\hat{O}=1/r$ or the damped Coulomb operator $\hat{O}=\textnormal{erfc}(\beta r)/r$ \cite{Jung2005}.
The most common is to use the overlap operator $\hat{O}=1$.
The minimization of eq. \eqref{metric} with respect to the expansion coefficients $c_k$ origin a linear system of equations:
\begin{equation}
    \frac{\partial E_\textnormal{self}}{\partial c_l} =
    -\sum_{\mu,\nu} P_{\mu\nu} \langle \mu\nu|\hat{O}|l \rangle + \sum_k c_k \langle k|\hat{O}|l \rangle
    \label{minimized_self_energy}
\end{equation}
where $P_{\mu\nu}$ is the density matrix.
The solution of Eq. \eqref{minimized_self_energy} requires the inversion of a the ABS matrix $\textbf{G}=\langle k|\hat{O}|l \rangle$.
Although in principle \textbf{G} must be positive definite and symmetric, in practice, it is almost singular and, therefore, the diagonalization to obtain its inverse must be done carefully.
To this end, both analytical and numerical approaches have
been developed to perform the fitting of the molecular densities
\cite{Cisneros2006,Cisneros2007}.

The latest GEM water model has been fitted to match the total energy
of the dimer surface and hexamer clusters reported by Paesani \etal, 
\cite{Babin2013} as well as the individual GEM terms (Coulomb,
exchange, polarization and Van der Waals) to reference symmetry 
adapted perturbation theory (SAPT) with 
aug--cc--pVTZ \cite{Duke2014}. The fitted density for GEM was calculated at the CCSD(T)/aug--cc--pVTZ level. \cite{Duke2014,Duke2019} The 
auxiliary basis sets employed  involve 70 primitives, resulting 
in approximately 280,000 functions  for a system of 512 waters. T
he integrals required for the calculation of the GEM
terms are evaluated using an extension of the sPME approach
for Gaussian functions, \cite{Duke2014,Duke2019}
which has been implemented in the 
gem.pmemd program, which is available with the AmberTools release \cite{AmberTools}.

\begin{figure}[H]
  \includegraphics[width=\textwidth]{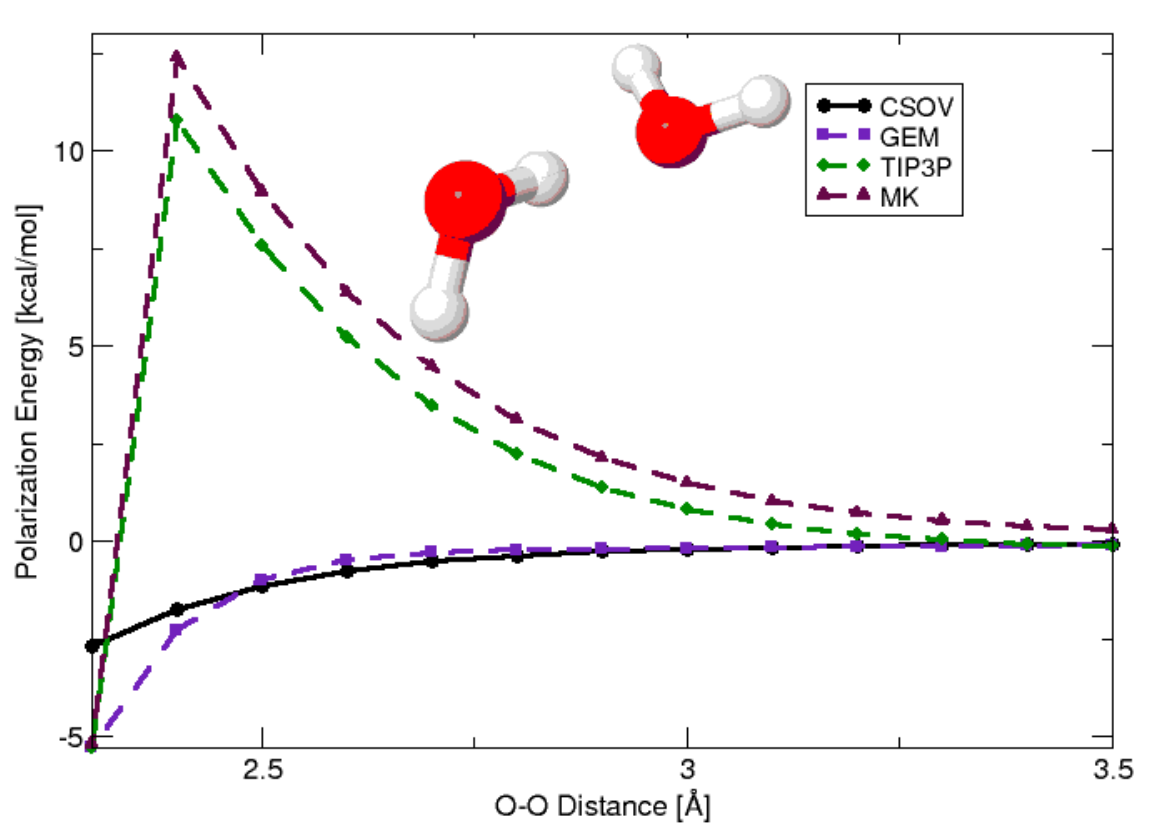}
  \caption{Polarization of the QM subsystem for one QM water by one GEM water. Adapted with permission from J. Phys. Chem. B. 2006, 110, 28, 13682–13684. Copyright 2006 American Chemical Society.}
  \label{fig:qm-gem-orig}
\end{figure}

The first implementation of the QM/GEM approach involved only
the Coulomb embedding term \cite{GEM2006-c}. The use of GEM
provides an accurate polarization environment for the QM wavefuntion
and  avoids penetration error effects because of the description of
the MM environment with explicit (frozen) densities compared with
conventional point charge force fields as shown in Figure 
\ref{fig:qm-gem-orig}.

For the current implementation, the total QM/MM interaction energy between the
subsystems, $E^{\textnormal{QM/GEM}}$, involves
four terms: Coulomb, Exchange--repulsion, polarization, and dispersion: $E_{\textnormal{Coul}}^{\textnormal{QM/GEM}}+E_{\textnormal{exch}}^{\textnormal{QM/GEM}}+E_{\textnormal{pol}}^{\textnormal{QM/GEM}}+E_{\textnormal{disp}}^{\textnormal{QM/GEM}}$ \cite{Lichem2019}.

The Coulomb and exchange--repulsion terms involve 3 center integrals between the QM
density and the fitted GEM density:
\begin{equation}
E_{\textnormal{Coul}}^{\textnormal{QM/GEM}}=\int\int\frac{\rho(r_1)\tilde{\rho}(r_2)}{r_{12}}dr_1dr_2
\end{equation}
and 
\begin{equation}
E_{\textnormal{exch}}^{\textnormal{QM/GEM}}=K_{\textnormal{exch}}\int\int\rho(r_1)\tilde{\rho}(r_2)dr_1dr_2,
\end{equation}
\noindent where $K_{\textnormal{exch}}$ is a proportionality constant \cite{Wheatly-Price-1990,Piquemal2006,Cisneros2006}.
The polarization interaction is calculated using the same approximation described above
for LICHEM with AMOEBA. The dispersion term 
is approximated by a multipolar expansion taking the 6, 8 and 12 terms into
consideration. The exchange proportionality parameter, as well as the coefficients
for the dispersion term have been parametrized by linear least squares to match
the SAPT2+3/aug--cc--pVTZ components for the ten water dimers \cite{Tschumper02690}.

\begin{figure}[t]
  \includegraphics[width=\textwidth]{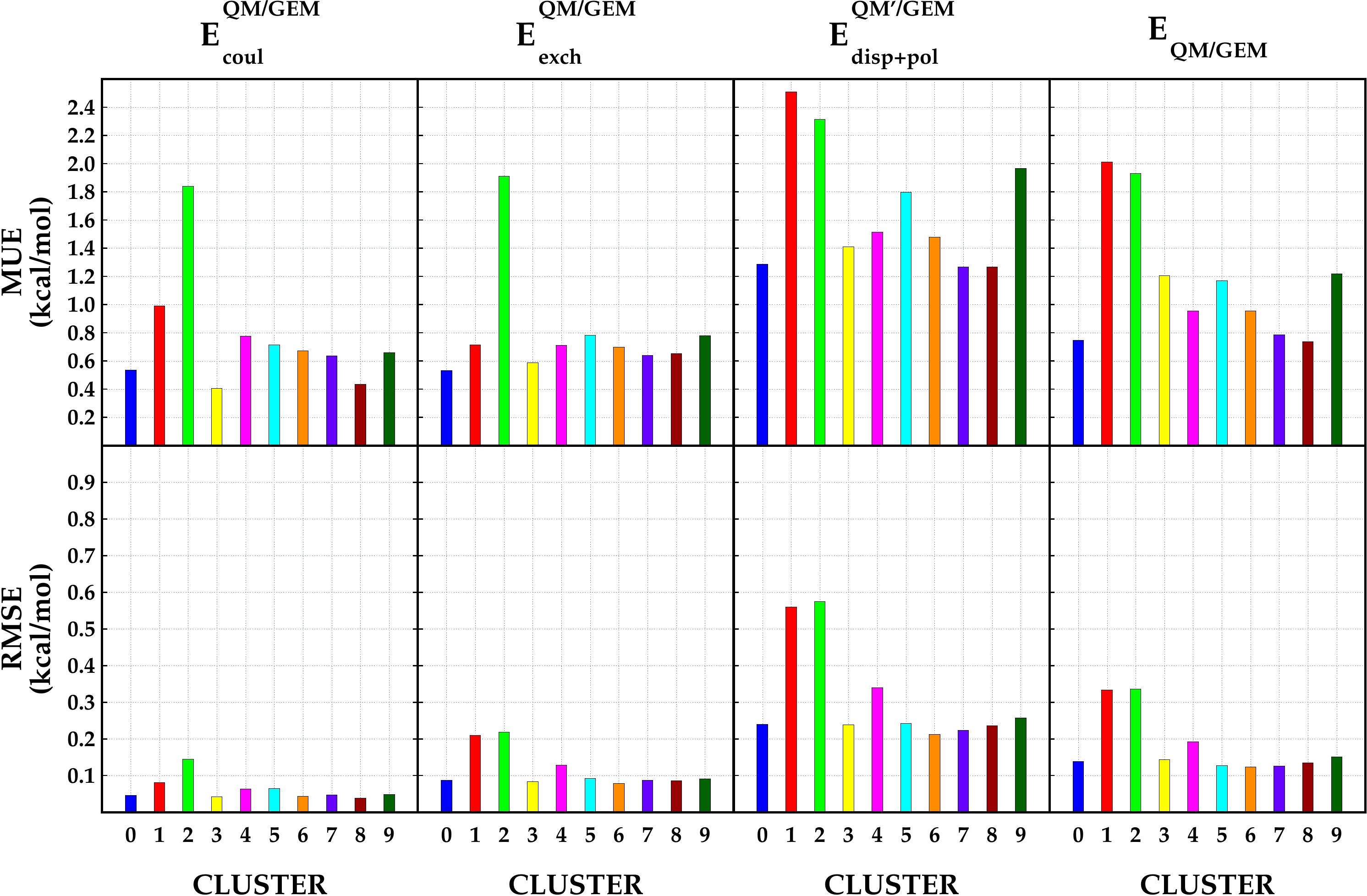}
  \caption{Errors per cluster with respect to SAPT. The errors corresponding to the Coulomb interaction energy are depicted on the first column, and the errors corresponding to the exchange interaction energy are given on the second column. The computed errors for the sum of dispersion and polarization energies are depicted on the third, while the error for total energy is given on the fourth column. Reprinted with permission from J. Phys. Chem. Lett. 2018, 9, 11, 3062–3067. Copyright 2018 American Chemical Society.}
  \label{fig:qm-gem}
\end{figure}

In this implementation we considered two approaches for the calculation
of the total interaction depending on the calculation of the QM/GEM exchange--repulsion
term. In both cases the QM/GEM dispersion
terms are added {\em a posteriori} to the total energy (and forces). The Coulomb
term is calculated by including the  GEM density in the core Hamiltonian in both
alternative approaches. Thus, the difference in the two approaches involves only
the exchange--repulsion component. In one approach, the exchange--repulsion inter--molecular
interaction is calculated after the SCF has completed
The second approach involves the inclusion of the exchange--repulsion term explicitly
in the SCF, i.e. explicit exchange-repulsion embedding. This is achieved by including the GEM density in the core Hamiltonian:
\begin{equation}
H_{\textnormal{eff}}=H_{\textnormal{core}}+V_{\textnormal{GEM}}
\end{equation}\label{eq.qmgem1}
\noindent in which $V_{GEM}$ involves the introduction of the 3--center integrals of the QM MOs with
the GEM density in the core Hamiltonian by:
\begin{equation}
V_{\textnormal{GEM}}= \sum_l x_l \sum_{\mu\nu} \langle \mu\nu\parallel l \rangle + K^{\prime}_{exch} \sum_l x_l \sum_{\mu\nu} \langle \mu\nu \mid l \rangle ,
\end{equation}\label{eq.qmgempotC}
\noindent where the first term corresponds to the Coulomb integrals and the second term to the
overlap integrals multiplied by the exchange--repulsion proportionality constant. In this
second case the exchange--repulsion proportionality constant is different than the former
case since the QM wavefunction experiences a different external potential (see below).
In all cases the required integrals have been programmed into a modified version of Psi4 \cite{psi4}.

The QM/GEM implementation in LICHEM was
assesed by comparing the total intermolecular
interaction Energies, as well as the
individual interaction terms for a sub--set
of the water dimer potential reported by
Babin {\em et al.} \cite{Babin2013} as
described in Ref. \cite{Lichem2019}. The
maximum unsigned error (MUE), and root mean 
squared error (RMSE) for the clustered dimers
is shown in Figure \ref{fig:qm-gem}.
In the original comparison, the term--by--term
comparison was done with respect to 
SAPT2+3/aug--cc--pVTZ, which does not
provide a straightforward separation of the
induction and polarization components. 
Therefore, the dispersion+polarization terms
were analyzed together. As can be
seen from Figure \ref{fig:qm-gem}, 
the use of GEM for the 
Coulomb, polarization and exchange--repulsion
embedding provides an accurate environment
for the hybrid calculation.

\section{LICHEM}
\label{sec:lichem}
Various methods described above including the pseudobond, and QM/MM--LREC, have been implemented in the LICHEM software (Layered Interacting CHEmical Models) package in the context of 
both non--polarizable and polarizable potentials including AMOEBA, GEM, and MB--Pol. \cite{Lichem2016, Lichem2019}.
At its core, LICHEM is an interface that can couple QM codes such as Gaussian \cite{Gaussian}, NWChem \cite{NWChem} and Psi4 \cite{psi4} and MM codes such as TINKER \cite{TINKER}, TINKER-HP \cite{TINKER-HP} and LAMMPS \cite{LAMMPS}.

When AMOEBA (or another multipolar force field) is employed, LICHEM uses
an approximation for the classical multipolar environment to allow the
seamless integration with multiple QM codes \cite{Devereux2014,Kratz2016}
LICHEM provides a variety of approaches for non--polarizable and
polarizable \qmmm simulations including single point energies, Monte
Carlo (Grand Canonical and Isothermal--Isobaric ensembles), and path integral
Monte Carlo. Geometry optimizations can be performed in an iterative procedure, where QM and MM atoms are optimized separately.
QM atoms are optimized including MM charges or multipoles.
Then, MM atoms are optimized including point charges comes from electron density from QM atoms.
This procedure continue until convergence criteria are met.



Geometry optimization algorithms for single structures are limited to
critical points along a surface. In many cases, it is useful to
determine the entire reaction path that connects two, or several 
critical points to investigate a reaction mechanism. An entire 
reaction path can be optimized by performing an optimization with 
multiple replicas between the reactant and product states. Several 
algorithms have been developed for this end. One such family is 
called the chain--of--replica methods. These methods represent a path
as a string of beads connected to each other. LICHEM has two 
chain--of--replica algorithms for path optimization, the Nudged 
Elastic Band (NEB) \cite{Jonsson1998,Henkelman2000}, and the
Quadratic String Method (QSM) \cite{Burger2006}.

The nudged elastic band (NEB) method \cite{Jonsson1998} is an efficient method for finding the minimum energy path (MEP) between a given initial and final state of a transition.
To sample regions between stationary points (minima or transition states), spring forces can be added to ensure continuity of the path.
In the NEB method, the modified forces ($F$') of the elastic band on the bead $p$ and the step $n$ are given by:
\begin{equation}
    F_{p,n}' =
    F_{p,n} + K \Delta R_{p,n}
\end{equation}
where $K$ is the spring constant and $\Delta R$ is the displacement between replicas defined as:
\begin{equation}
    \Delta R_{p,n} =
    |R_{p+1,n}-R_{p,n}|-|R_{p,n}-R_{p-1,n}|
\end{equation}

Spring forces contribute to the elastic band method in locating the minimum energy path.
However, to prevent the beads from slipping to the minimum, it is necessary to eliminate the forces parallel to the reaction.
The forces are then given by
\begin{equation}
    F_{p,n}' =
    F_{p,n} - F_{p,n}\cdot\tau_{p,n}\tau_{p,n} + K \Delta R_{p,n}\tau{p,n}
\end{equation}
where $\tau$ is a tangent unit vector indicating the direction of the reaction path.

The climbing image (CI) NEB method is a modification to the NEB method.
The IC travels upward along the tangent to find an approximate transition state structure. The transition state forces in the IC (replica c) are 
expressed as:
\begin{equation*}
    F_{c,n}' = F_{c,n}-2F_{c,n} \cdot \tau_{c,n}\tau_{c,n}
\end{equation*}
where the forces are inverted along the tangent \cite{Henkelman2000}.


\begin{figure}[t!]
    \centering
    \includegraphics[width=1\textwidth]{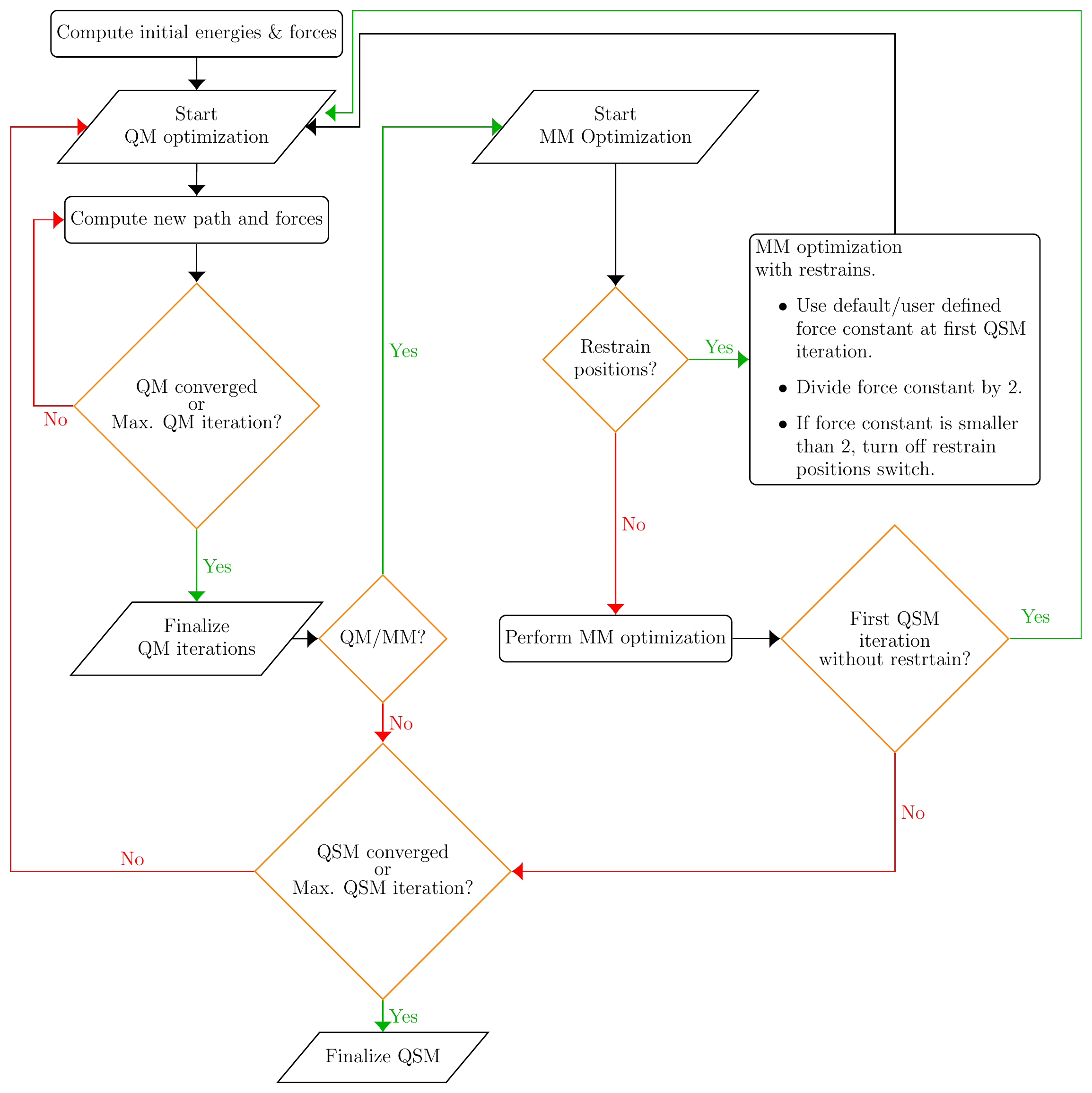}
    \caption{Flowchart of QSM algorithm implemented within LICHEM.
    Reprinted with permission from Hatice Gökcan, Erik Antonio Vázquez-Montelongo, and G. Andrés Cisneros. Journal of Chemical Theory and Computation 2019 15 (5), 3056-3065. Copyright (2020) American Chemical Society.}
    \label{fig:optflow}
\end{figure}

The Quadratic String Method (QSM) \cite{Burger2006} does not require 
a predetermined step size and eliminates the need for an artificial 
spring constant. Instead, the approximate Hessian confidence radius 
and surface are updated at the end of each iteration.
First, the QSM algorithm calculates the energy and gradients of all 
user-defined beads. The approximate Hessians ($H_k$) constructed from
the gradients in the initial step are updated with the damped 
Broyden-Fletcher-Goldfarb-Shanno (DBFGS) algorithm \cite{Fletcher1987}

\begin{equation}
    H^{i+1} = H^i -
    \frac{H^i \delta^i (\delta^i)^T H^i}{(\delta^i)^T H^i \delta^i} +
    \frac{r^i(r^i)^T}{(r^i)^T\delta^i}
\end{equation}
where $i$ indicates to the iteration number and
\begin{equation}
    \delta^i =
    x^{i+1}-x^i
\end{equation}
\begin{equation}
    r^i = \theta^i \gamma^i + (1-\theta^i) H^i \delta^i 
\end{equation}
with
\begin{equation}
    y^i = g^{i+1}-g^i
\end{equation}
\begin{equation}
    \theta^i =
    \begin{cases}
    1, \quad \text{if} \quad (\delta^i)^T \gamma^i > 0.2 (\delta^i)^T H^i \delta^i\\
    \displaystyle \frac{0.8(\delta^i)^T H^i\delta^i}{(\delta^i)^T H^k\delta^k - (\delta^i)^T}, \quad \text{otherwise}
\end{cases}
\end{equation}
The update of $H_k$ is followed by the update of the trust radii using the energy as a merit function;
\begin{equation}
    \rho = \frac{E_k^{i+1}-E_k^i}{dx_k^Tg_k^i+ \displaystyle \frac{1}{2}dx_k^T H_k^i dx_k}
\end{equation}


\begin{figure}[t!]
    \centering
    \includegraphics[width=0.9\textwidth]{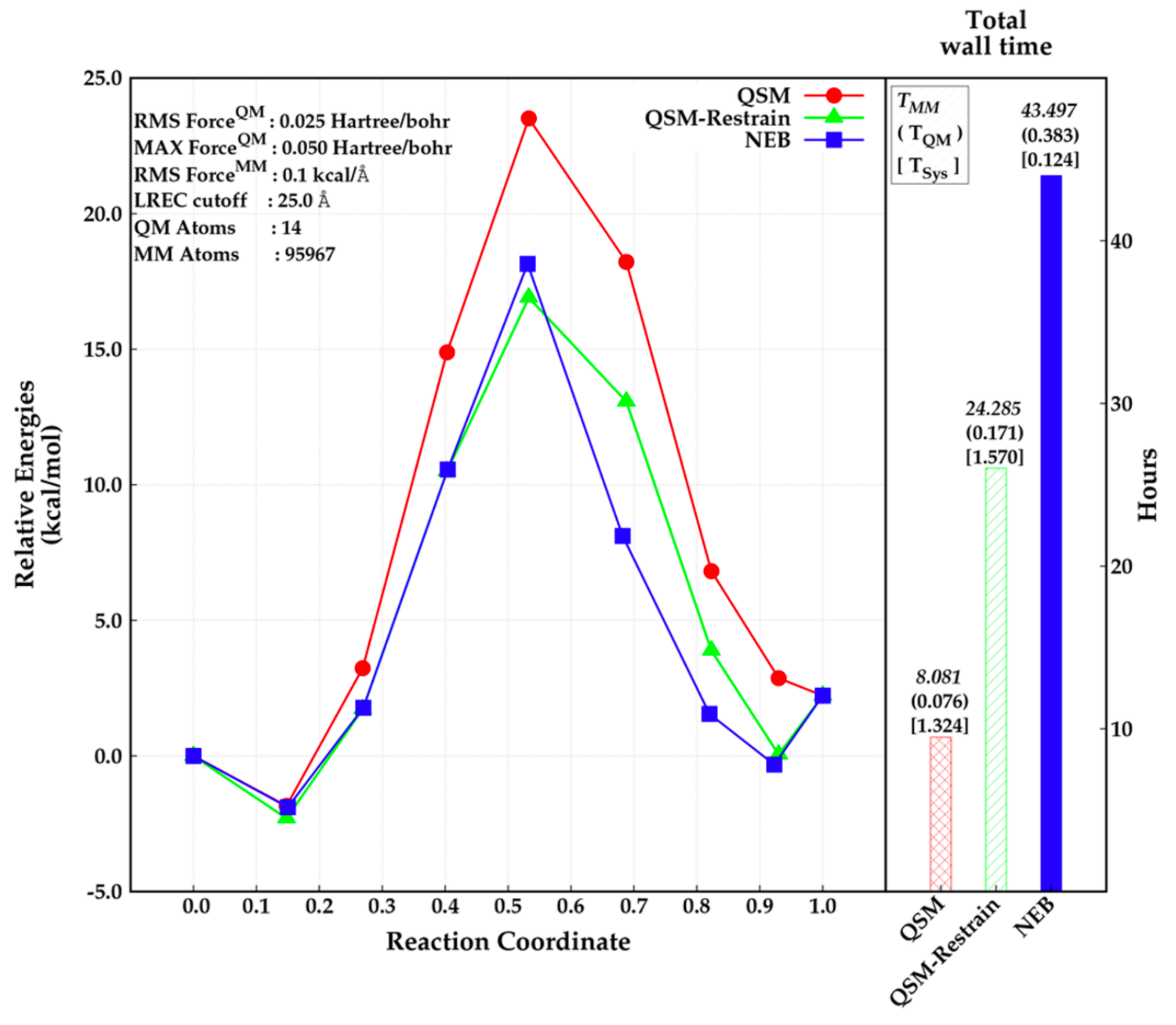}
    \caption{Relative energies obtained by NEB (blue, square) and QSM (red, circle), and QSM using the restrained-MM method (green, triangle) along the reaction coordinate (left), and total wall time required for the optimization in hours (right). The wall time for MM calculations, QM calculations, and for system calls are depicted above the wall time bars (right). Reprinted with permission from Hatice Gökcan, Erik Antonio Vázquez-Montelongo, and G. Andrés Cisneros. Journal of Chemical Theory and Computation 2019 15 (5), 3056-3065. Copyright (2019) American Chemical Society}.
    \label{fig:neb-qsm}
\end{figure}

Both NEB and QSM require optimized structures for the end--points
(e.g. reactant and product) as well as an initial guess of the intermediate images
(beads). One approach to generate this initial guess is to use a
linear interpolation between the end--point coordinates. In \qmmm 
simulations this is usually accomplished using only the coordinates
of the QM subsystem, and use the MM environment from one of the ends
for all the beads. However, this can
leads to issues during the path optimization such as
discontinuities along the trajectory.
One approach to avoid this is to employ a restraints MM optimization
approach \cite{Xie2004,Cisneros2005b}.

In this case, the initial stages of the iterative QM/MM
path optimization are performed with restraints on the MM atoms during the MM part of the optimization.
These restraints are reduced at each subsequent iteration. \figrff{fig:optflow} shows a flowchart that describes these approaches in LICHEM.


One more algorithmic advantage that can be applied in the case of
"chain--of--replica" approaches is the fact that each bead 
can be calculated independently. Therefore, it is possible to employ 
hybrid-parallelization paradigms such that the QM or MM parts of the
calculation employ the parallelism inherent in the respective codes
(i.e. MPI, OpenMP, cuda, etc) and the path optimizer used MPI to 
run all beads in parallel. This approach has been implemented in 
LICHEM for QSM \cite{Gokcan2018}.
\figrff{fig:neb-qsm} shows
performance metrics for a test reaction comparing NEB, QSM and 
QSM with restrained MM optimization.

One example of the use of the methods mentioned above for a QM/MM calculation with AMOEBA was reported by V\'azquez-Montelongo \etal \cite{Vazquez-Montelongo2018}, who performed polarizable \qmmm simulations to investigate the \textit{N-tert}-Butyl\-oxycarbonylation (Boc) of aniline in ionic liquid mixture composed of water/ [EMIm][BF$_4$]. 
Boc is a common group used to protect molecules in organic synthesis \cite{Sarkar2011}. 

\begin{figure}[H]

\begin{center}

\subfloat[]{\includegraphics[width=\textwidth]{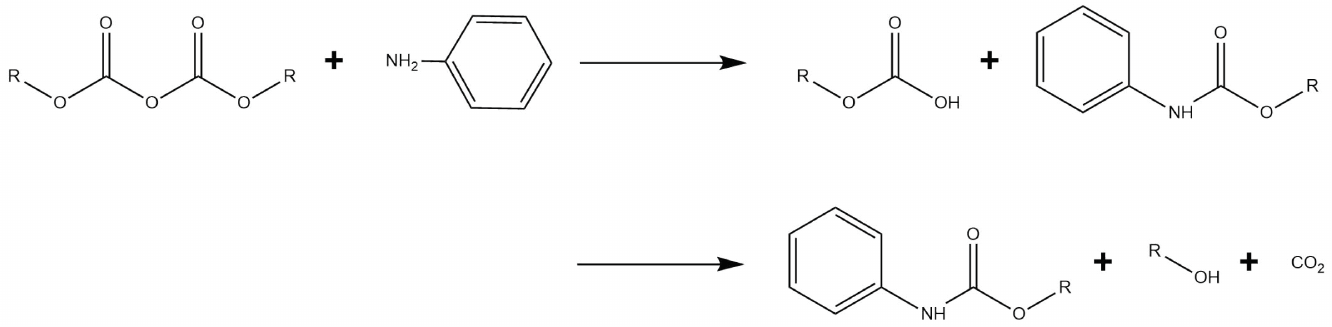}
\label{fig:aniline-reaction-c}}

\subfloat[]{\includegraphics[width=.5\textwidth]{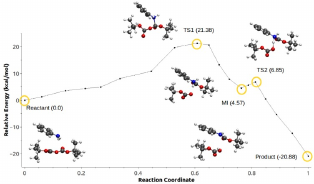}
\label{aniline-1c-pol}}
\subfloat[]{\includegraphics[width=.5\textwidth]{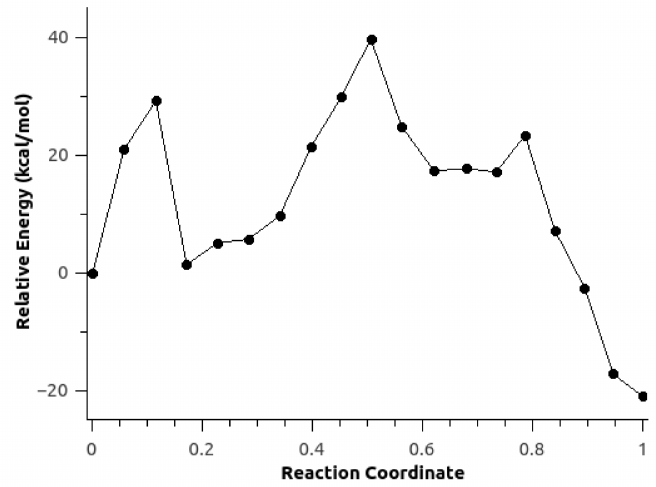}
\label{aniline-1c-no-pol}}

\subfloat[]{\includegraphics[width=\textwidth]{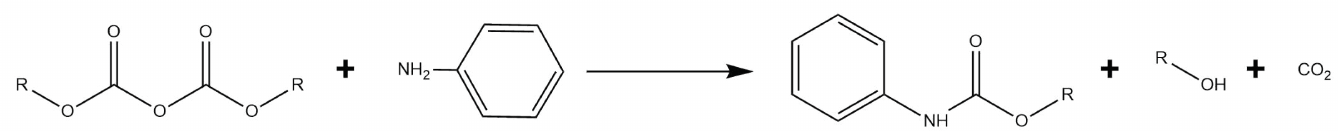}
\label{fig:aniline-reaction-d}}

\subfloat[]{\includegraphics[width=.5\textwidth]{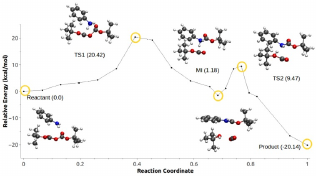}
\label{aniline-1d-pol}}
\subfloat[]{\includegraphics[width=.5\textwidth]{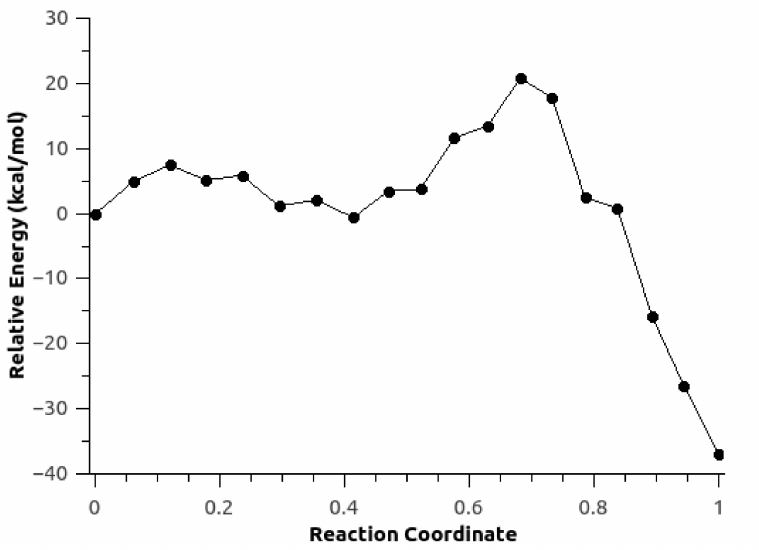}
\label{aniline-1d-no-pol}}

\caption{\small
\protect\subref{fig:aniline-reaction-c} Reaction scheme for the N-tert-butoxycarbonylation of aniline for the step–wise mechanism. 
\protect\subref{aniline-1c-pol} Minimum energy path for Configuration C2, Scheme 1c.
\protect\subref{aniline-1c-no-pol} Minimum energy path for Scheme c for Configuration C2 without the AMOEBA polarization term.
\protect\subref{fig:aniline-reaction-d} Reaction scheme for the N-tert-butoxycarbonylation of aniline for the concerted mechanism.
\protect\subref{aniline-1d-pol} Minimum energy path for Configuration C2, Scheme 1d. 
\protect\subref{aniline-1d-no-pol} Minimum energy path for Scheme d for Configuration C2 without the AMOEBA polarization term.
Adapted from Vazquez-Montelongo, E.A.; Vazquez-Cervantes, J.E.; Cisneros, G.A. Polarizable ab initio QM/MM Study of the Reaction Mechanism of N-tert-Butyloxycarbonylation of Aniline in [EMIm][BF4]. Molecules 2018, 23, 2830. \cite{Vazquez-Montelongo2018}} 
\label{fig:aniline-path}
\end{center}
\end{figure}

V\'azquez-Montelongo \etal have assessed the role of the solvent in four different possible reaction mechanisms of \textit{N-tert}-Boc protection of aniline using \lichem. Two of them require an activation of the Boc group by a water molecule or a cation ([EMIm]$^+$), while the other two do not (denoted configuration C2). Here we will focus on the two possible mechanisms of configuration C2 which are shown in Figures \ref{fig:aniline-reaction-c} and \ref{fig:aniline-reaction-d}. Reaction (1) (Fig. 
\ref{fig:aniline-reaction-c}, denoted scheme 1c in the
original study) represents a sequential mechanism: first, the nucleophic attack of the aniline to the carbonyl of one of the Boc groups occurs, followed by the formation of CO\textsubscript{2}.
Reaction (2) corresponds to the concerted mechanism (Fig. 
\ref{fig:aniline-reaction-d}, denoted scheme 1d in 
the original study) where the nucleophic attack of the aniline to the carbonyl of one of the Boc groups and the formation of CO\textsubscript{2} occur concurrently. 


The path optimization methods described above have been applied to obtain minimum energy paths (MEP) for both  reaction schemes (Figures \ref{aniline-1c-pol} and \ref{aniline-1d-pol}). 
The calculated rate--limiting barrier for reaction (1) is 21.38 kcal/mol
compared with 20.42 kcal/mol for reaction (2). In order to assess the 
effect of polarization on reaction paths, the MEPs were re-computed with the AMOEBA polarization term set to zero during the calculation (Figures \ref{aniline-1c-no-pol} and \ref{aniline-1d-no-pol}).
In this case, MEP with or without the polarization terms give similar barriers for  reaction (1), although the shape of the reaction path
without polarization is drastically altered, and the intermediate
and TS2 observed in the original path are no longer observed. 
In the case of reaction (2) the differences between the calculated paths with and without polarization are more significant. The energy barrier computed between the reactant and product increases significantly when the polarization is not taken into account, and the shape of the path is also affected. This example illustrates the necessity to include explicit polarization within \qmmm calculations in highly charged environments.

\newpage

\section{Conclusions}
\label{sec:conclu}
The computational study of reactions in biological and condensed-phase
systems has been advanced the combination of \qm and \mm methods.
Here, we have reported recent advances in the development of \qmmm
including polarizable force fields. The use of polarizable
potentials such as AMOEBA, gives rise to novel issues that need to be
considered, including challenges at the internal and external
boundaries (long--range effects) and coupling to the QM 
software. This review focused on the methods and approaches
that have been developed and implemented in LICHEM to address
these, and other challenges, such as the extension of the pseudobond
approach, the development and implementation of QM/MM--LREc, 
the implementation of restrained MM path optimizations coupled with
the quadratic string method, implementation of advanced force
fields for \qmmm such as MB--Pol and GEM, etc. Several examples 
have been provided to underscore the utility of polarizable 
potentials and their usefulness in the study of complex 
biological and condensed phase systems.



\bibliographystyle{plain}
\bibliography{ms}

\begin{thebibliography}{100}

\bibitem{Acevedo2014}
Orlando Acevedo.
\newblock {Simulating Chemical Reactions in Ionic Liquids Using QM/MM
  Methodology}.
\newblock {\em The Journal of Physical Chemistry A}, 118(50):11653--11666, dec
  2014.

\bibitem{Acevedo2014b}
Orlando Acevedo and Wiliiam~L Jorgensen.
\newblock {Quantum and molecular mechanical Monte Carlo techniques for modeling
  condensed-phase reactions}.
\newblock {\em WIREs Computational Molecular Science}, 4(5):422--435, sep 2014.

\bibitem{Ahmadi2018}
Shideh Ahmadi, Lizandra {Barrios Herrera}, Morteza Chehelamirani, Jiř{\'{i}}
  Hosta{\v{s}}, Said Jalife, and Dennis~R Salahub.
\newblock {Multiscale modeling of enzymes: QM-cluster, QM/MM, and QM/MM/MD: A
  tutorial review}.
\newblock {\em International Journal of Quantum Chemistry}, 118(9):e25558,
  2018.

\bibitem{Allinger1996}
Norman~L Allinger, Kuohsiang Chen, and Jenn-Huei Lii.
\newblock {An improved force field (MM4) for saturated hydrocarbons}.
\newblock {\em Journal of Computational Chemistry}, 17(5‐6):642--668, apr
  1996.

\bibitem{Allinger1989}
Norman~L Allinger, Young~H Yuh, and Jenn~Huei Lii.
\newblock {Molecular mechanics. The MM3 force field for hydrocarbons. 1}.
\newblock {\em Journal of the American Chemical Society}, 111(23):8551--8566,
  nov 1989.

\bibitem{Amara2000}
Patricia Amara, Martin~J. Field, Cristobal Alhambra, and Jiali Gao.
\newblock {The generalized hybrid orbital method for combined quantum
  mechanical/molecular mechanical calculations: formulation and tests of the
  analytical derivatives}.
\newblock {\em Theoretical Chemistry Accounts: Theory, Computation, and
  Modeling (Theoretica Chimica Acta)}, 104(5):336--343, aug 2000.

\bibitem{Aqvist1993}
Johan {\AA}qvist and Arieh Warshel.
\newblock {Simulation of enzyme reactions using valence bond force fields and
  other hybrid quantum/classical approaches}.
\newblock {\em Chemical Reviews}, 93(7):2523--2544, nov 1993.

\bibitem{Assfeld1996}
Xavier Assfeld and Jean-Louis Rivail.
\newblock {Quantum chemical computations on parts of large molecules: the ab
  initio local self consistent field method}.
\newblock {\em Chemical Physics Letters}, 263(1-2):100--106, dec 1996.

\bibitem{Babin2013}
Volodymyr Babin, Claude Leforestier, and Francesco Paesani.
\newblock {Development of a “First Principles” Water Potential with
  Flexible Monomers: Dimer Potential Energy Surface, VRT Spectrum, and Second
  Virial Coefficient}.
\newblock {\em Journal of Chemical Theory and Computation}, 9(12):5395--5403,
  2013.

\bibitem{Beglov1994}
Dmitrii Beglov and Beno{\^{i}}t Roux.
\newblock {Finite representation of an infinite bulk system: Solvent boundary
  potential for computer simulations}.
\newblock {\em The Journal of Chemical Physics}, 100(12):9050--9063, 1994.

\bibitem{Benighaus2011}
Tobias Benighaus and Walter Thiel.
\newblock {Long-Range Electrostatic Effects in QM/MM Studies of Enzymatic
  Reactions: Application of the Solvated Macromolecule Boundary Potential}.
\newblock {\em Journal of Chemical Theory and Computation}, 7(1):238--249, jan
  2011.

\bibitem{Bondanza2020}
Mattia Bondanza, Michele Nottoli, Lorenzo Cupellini, Filippo Lipparini, and
  Benedetta Mennucci.
\newblock {Polarizable embedding QM/MM: the future gold standard for complex
  (bio)systems?}
\newblock {\em Phys. Chem. Chem. Phys.}, 22(26):14433--14448, 2020.

\bibitem{Boulanger2018}
Eliot Boulanger, Lei Huang, Chetan Rupakheti, Alexander~D {MacKerell Jr}, and
  Beno{\^{i}}t Roux.
\newblock {Optimized Lennard-Jones Parameters for Druglike Small Molecules}.
\newblock {\em Journal of chemical theory and computation}, 14(6):3121--3131,
  jun 2018.

\bibitem{Boulanger2012}
Eliot Boulanger and Walter Thiel.
\newblock {Solvent Boundary Potentials for Hybrid QM/MM Computations Using
  Classical Drude Oscillators: A Fully Polarizable Model}.
\newblock {\em Journal of Chemical Theory and Computation}, 8(11):4527--4538,
  nov 2012.

\bibitem{Brooks1983}
Bernard~R Brooks, Robert~E Bruccoleri, Barry~D Olafson, David~J States,
  S~Swaminathan, and Martin Karplus.
\newblock {CHARMM: A program for macromolecular energy, minimization, and
  dynamics calculations}.
\newblock {\em Journal of Computational Chemistry}, 4(2):187--217, 1983.

\bibitem{Brunk2015}
Elizabeth Brunk and Ursula Rothlisberger.
\newblock {Mixed Quantum Mechanical/Molecular Mechanical Molecular Dynamics
  Simulations of Biological Systems in Ground and Electronically Excited
  States}.
\newblock {\em Chemical Reviews}, 115(12):6217--6263, jun 2015.

\bibitem{Burger2006}
Steven~K Burger and Weitao Yang.
\newblock {Quadratic string method for determining the minimum-energy path
  based on multiobjective optimization}.
\newblock {\em The Journal of Chemical Physics}, 124(5):54109, 2006.

\bibitem{AmberTools}
D.A. Case, K.~Belfon, I.Y. Ben-Shalom, S.R. Brozell, D.S. Cerutti, T.E.
  Cheatham, V.W.D. Cruzeiro, T.A. Darden, R.E. Duke, G.~Giambasu, M.K. Gilson,
  H.~Gohlke, A.W. Goetz, R.~Harris, S.~Izadi, S.A. Izmailov, K.~Kasavajhala,
  A.~Kovalenko, R.~Krasny, T.~Kurtzman, T.S. Lee, S.~LeGrand, P.~Li, C.~Lin,
  J.~Liu, T.~Luchko, R.~Luo, V.~Man, K.M. Merz, Y.~Miao, O.~Mikhailovskii,
  G.~Monard, H.~Nguyen, A.~Onufriev, F.~Pan, S.~Pantano, R.~Qi, D.R. Roe,
  A.~Roitberg, C.~Sagui, S.~Schott-Verdugo, J.~Shen, C.L. Simmerling,
  N.R.Skrynnikov, J.~Smith, J.~Swails, R.C. Walker, J.~Wang, L.~Wilson, R.M.
  Wolf, X.~Wu, Y.~Xiong, Y.~Xue, D.M. York, and P.A. Kollman.
\newblock {AMBER 2020}.
\newblock {\em University of California, San Francisco.}, 2020.

\bibitem{Case2005}
David~A Case, Thomas~E {Cheatham III}, Tom Darden, Holger Gohlke, Ray Luo,
  Kenneth~M {Merz Jr.}, Alexey Onufriev, Carlos Simmerling, Bing Wang, and
  Robert~J Woods.
\newblock {The Amber biomolecular simulation programs}.
\newblock {\em Journal of Computational Chemistry}, 26(16):1668--1688, 2005.

\bibitem{Chen2020}
Xin Chen and Jiali Gao.
\newblock {Fragment Exchange Potential for Realizing Pauli Deformation of
  Interfragment Interactions}.
\newblock {\em The Journal of Physical Chemistry Letters}, 11(10):4008--4016,
  may 2020.

\bibitem{Chen40082020}
Xin Chen, Zexing Qu, Bingbing Suo, and Jiali Gao.
\newblock {A self-consistent coulomb bath model using density fitting}.
\newblock {\em Journal of Computational Chemistry}, 41(18):1698--1708, 2020.

\bibitem{Cisneros2007}
G~Andr{\'{e}}s Cisneros, Dennis Elking, Jean-Philip Piquemal, and Thomas~A
  Darden.
\newblock {Numerical fitting of molecular properties to Hermite Gaussians}.
\newblock {\em The journal of physical chemistry. A}, 111(47):12049--12056, nov
  2007.

\bibitem{Cisneros2005b}
G~Andr{\'{e}}s Cisneros, Haiyan Liu, Zhenyu Lu, and Weitao Yang.
\newblock {Reaction path determination for quantum mechanical/molecular
  mechanical modeling of enzyme reactions by combining first order and second
  order "chain-of-replicas" methods.}
\newblock {\em The Journal of chemical physics}, 122(11):114502, mar 2005.

\bibitem{Cisneros2005}
G~Andr{\'{e}}s Cisneros, Jean-Philip Piquemal, and Thomas~A Darden.
\newblock {Intermolecular electrostatic energies using density fitting}.
\newblock {\em The Journal of Chemical Physics}, 123(4):44109, 2005.

\bibitem{Cisneros2006}
G~Andr{\'{e}}s Cisneros, Jean-Philip Piquemal, and Thomas~A Darden.
\newblock {Generalization of the Gaussian electrostatic model: Extension to
  arbitrary angular momentum, distributed multipoles, and speedup with
  reciprocal space methods}.
\newblock {\em The Journal of Chemical Physics}, 125(18):184101, 2006.

\bibitem{GEM2006-c}
G~Andr{\'{e}}s Cisneros, Jean-Philip Piquemal, and Thomas~A Darden.
\newblock {Quantum Mechanics/Molecular Mechanics Electrostatic Embedding with
  Continuous and Discrete Functions}.
\newblock {\em The Journal of Physical Chemistry B}, 110(28):13682--13684, jul
  2006.

\bibitem{Darden2010}
Thomas~A Darden.
\newblock {Extensions of the Ewald method for Coulomb interactions in
  crystals}.
\newblock {\em International Tables for Crystallography}, pages 458--481, jun
  2010.

\bibitem{Das2002}
Debananda Das, Kirsten~P. Eurenius, Eric~M. Billings, Paul Sherwood, David~C.
  Chatfield, Milan Hodo{\v{s}}{\v{c}}ek, and Bernard~R. Brooks.
\newblock {Optimization of quantum mechanical molecular mechanical partitioning
  schemes: Gaussian delocalization of molecular mechanical charges and the
  double link atom method}.
\newblock {\em The Journal of Chemical Physics}, 117(23):10534--10547, dec
  2002.

\bibitem{DasR2019}
Ranjita Das, Erik~A V{\'{a}}zquezv{\'{a}}zquez-Montelongo, G~Andr{\'{e}}s,
  Andr{\'{e}}s Cisneros, and Judy~I Wu.
\newblock {Ground State Destabilization in Uracil DNA Glycosylase: Let's Not
  Forget "Tautomeric Strain" in Substrates}.
\newblock {\em Journal of the American Chemical Society}, 2019.

\bibitem{Devereux2014}
Mike Devereux, Shampa Raghunathan, Dmitri~G Fedorov, and Markus Meuwly.
\newblock {A Novel, Computationally Efficient Multipolar Model Employing
  Distributed Charges for Molecular Dynamics Simulations}.
\newblock {\em Journal of Chemical Theory and Computation}, 10(10):4229--4241,
  oct 2014.

\bibitem{DiLabio2002}
Gino~A. DiLabio, Margaret~M. Hurley, and Phillip~A. Christiansen.
\newblock {Simple one-electron quantum capping potentials for use in hybrid
  QM/MM studies of biological molecules}.
\newblock {\em The Journal of Chemical Physics}, 116(22):9578--9584, jun 2002.

\bibitem{Duarte2015}
Fernanda Duarte, Beat~A Amrein, David Blaha-Nelson, and Shina C~L Kamerlin.
\newblock {Recent advances in QM/MM free energy calculations using reference
  potentials}.
\newblock {\em Biochimica et biophysica acta}, 1850(5):954--965, may 2015.

\bibitem{Duke2019}
Robert~E Duke and G~Andr{\'{e}}s Cisneros.
\newblock {Ewald-based methods for Gaussian integral evaluation: application to
  a new parameterization of GEM}.
\newblock {\em Journal of molecular modeling}, 25(10):307, sep 2019.

\bibitem{Duke2014}
Robert~E Duke, Oleg~N Starovoytov, Jean-Philip Piquemal, and G~Andr{\'{e}}s
  Cisneros.
\newblock {GEM*: A Molecular Electronic Density-Based Force Field for Molecular
  Dynamics Simulations}.
\newblock {\em Journal of Chemical Theory and Computation}, 10(4):1361--1365,
  apr 2014.

\bibitem{Duster2018}
Adam Duster, Chun-Hung Wang, and Hai Lin.
\newblock {Adaptive QM/MM for Molecular Dynamics Simulations: 5. On the
  Energy-Conserved Permuted Adaptive-Partitioning Schemes}.
\newblock {\em Molecules}, 23(9):2170, aug 2018.

\bibitem{Dziedzic2019}
Jacek Dziedzic, Teresa Head-Gordon, Martin Head-Gordon, and Chris-Kriton
  Skylaris.
\newblock {Mutually polarizable QM/MM model with {\textless}i{\textgreater}
  {\textless}b{\textgreater}in situ{\textless}/b{\textgreater}
  {\textless}/i{\textgreater} optimized localized basis functions}.
\newblock {\em The Journal of Chemical Physics}, 150(7):074103, feb 2019.

\bibitem{Dziedzic2016}
Jacek Dziedzic, Yuezhi Mao, Yihan Shao, Jay Ponder, Teresa Head-Gordon, Martin
  Head-Gordon, and Chris-Kriton Skylaris.
\newblock {TINKTEP: A fully self-consistent, mutually polarizable QM/MM
  approach based on the AMOEBA force field}.
\newblock {\em The Journal of Chemical Physics}, 145(12):124106, sep 2016.

\bibitem{Essmann1995}
Ulrich Essmann, Lalith Perera, Max~L Berkowitz, Tom Darden, Hsing Lee, and
  Lee~G Pedersen.
\newblock {A smooth particle mesh Ewald method}.
\newblock {\em The Journal of Chemical Physics}, 103(19):8577--8593, nov 1995.

\bibitem{Ewald1921}
P~P Ewald.
\newblock {Die Berechnung optischer und elektrostatischer Gitterpotentiale}.
\newblock {\em Annalen der Physik}, 369(3):253--287, jan 1921.

\bibitem{Fang2015}
Dong Fang, Robert~E Duke, and G~Andr{\'{e}}s Cisneros.
\newblock {A new smoothing function to introduce long-range electrostatic
  effects in QM/MM calculations}.
\newblock {\em The Journal of Chemical Physics}, 143(4):44103, 2015.

\bibitem{Fang2013}
Dong Fang, Richard~L Lord, and G~Andr{\'{e}}s Cisneros.
\newblock {Ab Initio QM/MM Calculations Show an Intersystem Crossing in the
  Hydrogen Abstraction Step in Dealkylation Catalyzed by AlkB}.
\newblock {\em The Journal of Physical Chemistry B}, 117(21):6410--6420, may
  2013.

\bibitem{Ferenczy1992}
Gy{\"{o}}rgy~G. Ferenczy, Jean-Louis Rivail, P{\'{e}}ter~R. Surj{\'{a}}n, and
  G{\'{a}}bor N{\'{a}}ray-Szab{\'{o}}.
\newblock {NDDO fragment self-consistent field approximation for large
  electronic systems}.
\newblock {\em Journal of Computational Chemistry}, 13(7):830--837, sep 1992.

\bibitem{Ferre2002}
Nicolas Ferr{\'{e}}, Xavier Assfeld, and Jean-Louis Rivail.
\newblock {Specific force field parameters determination for the hybrid
  {\textless}i{\textgreater}ab initio{\textless}/i{\textgreater} QM/MM LSCF
  method}.
\newblock {\em Journal of Computational Chemistry}, 23(6):610--624, apr 2002.

\bibitem{Field1990}
Martin~J Field, Paul~A Bash, and Martin Karplus.
\newblock {A combined quantum mechanical and molecular mechanical potential for
  molecular dynamics simulations}.
\newblock {\em Journal of Computational Chemistry}, 11(6):700--733, 1990.

\bibitem{Fletcher1987}
R~Fletcher.
\newblock {\em {Practical Methods of Optimization; (2nd Ed.)}}.
\newblock Wiley-Interscience, USA, 1987.

\bibitem{Gaussian}
M~J Frisch, G~W Trucks, H~B Schlegel, G~E Scuseria, M~A Robb, J~R Cheeseman,
  G~Scalmani, V~Barone, G~A Petersson, H~Nakatsuji, X~Li, M~Caricato, A~V
  Marenich, J~Bloino, B~G Janesko, R~Gomperts, B~Mennucci, H~P Hratchian, J~V
  Ortiz, A~F Izmaylov, J~L Sonnenberg, D~Williams-Young, F~Ding, F~Lipparini,
  F~Egidi, J~Goings, B~Peng, A~Petrone, T~Henderson, D~Ranasinghe, V~G
  Zakrzewski, J~Gao, N~Rega, G~Zheng, W~Liang, M~Hada, M~Ehara, K~Toyota,
  R~Fukuda, J~Hasegawa, M~Ishida, T~Nakajima, Y~Honda, O~Kitao, H~Nakai,
  T~Vreven, K~Throssell, J~A {Montgomery Jr.}, J~E Peralta, F~Ogliaro, M~J
  Bearpark, J~J Heyd, E~N Brothers, K~N Kudin, V~N Staroverov, T~A Keith,
  R~Kobayashi, J~Normand, K~Raghavachari, A~P Rendell, J~C Burant, S~S Iyengar,
  J~Tomasi, M~Cossi, J~M Millam, M~Klene, C~Adamo, R~Cammi, J~W Ochterski, R~L
  Martin, K~Morokuma, O~Farkas, J~B Foresman, and D~J Fox.
\newblock {Gaussian 16 Revision C.01}.
\newblock {\em Gaussian Inc. Wallingford CT}, 2016.

\bibitem{JialiGao1998}
Jiali Gao, Patricia Amara, Cristobal Alhambra, and Martin~J. Field.
\newblock {A Generalized Hybrid Orbital (GHO) Method for the Treatment of
  Boundary Atoms in Combined QM/MM Calculations}.
\newblock {\em The Journal of Physical Chemistry A}, 1998.

\bibitem{Geerke2007}
Daan~P Geerke and Wilfred~F van Gunsteren.
\newblock {On the Calculation of Atomic Forces in Classical Simulation Using
  the Charge-on-Spring Method To Explicitly Treat Electronic Polarization}.
\newblock {\em Journal of Chemical Theory and Computation}, 3(6):2128--2137,
  nov 2007.

\bibitem{Giese2016}
Timothy~J Giese and Darrin~M York.
\newblock {Ambient-Potential Composite Ewald Method for ab Initio Quantum
  Mechanical/Molecular Mechanical Molecular Dynamics Simulation}.
\newblock {\em Journal of Chemical Theory and Computation}, 12(6):2611--2632,
  jun 2016.

\bibitem{Giese3830022017}
Timothy~J Giese and Darrin~M York.
\newblock {Quantum mechanical force fields for condensed phase molecular
  simulations.}
\newblock {\em Journal of physics. Condensed matter : an Institute of Physics
  journal}, 29(38):383002, sep 2017.

\bibitem{Giovannini2019}
Tommaso Giovannini, Alessandra Puglisi, Matteo Ambrosetti, and Chiara Cappelli.
\newblock {Polarizable QM/MM Approach with Fluctuating Charges and Fluctuating
  Dipoles: The QM/FQF$\mu$ Model}.
\newblock {\em Journal of Chemical Theory and Computation}, 15(4):2233--2245,
  apr 2019.

\bibitem{Giovannini2019b}
Tommaso Giovannini, Rosario~Roberto Riso, Matteo Ambrosetti, Alessandra
  Puglisi, and Chiara Cappelli.
\newblock {Electronic transitions for a fully polarizable QM/MM approach based
  on fluctuating charges and fluctuating dipoles: Linear and corrected linear
  response regimes}.
\newblock {\em The Journal of Chemical Physics}, 151(17):174104, nov 2019.

\bibitem{Gokcan2018}
Hatice G{\"{o}}kcan, Eric Kratz, Thomas~A Darden, Jean-Philip Piquemal, and
  G~Andr{\'{e}}s Cisneros.
\newblock {QM/MM Simulations with the Gaussian Electrostatic Model: A
  Density-based Polarizable Potential}.
\newblock {\em The Journal of Physical Chemistry Letters}, 9(11):3062--3067,
  jun 2018.

\bibitem{Lichem2019}
Hatice G{\"{o}}kcan, Erik~Antonio V{\'{a}}zquez-Montelongo, and G.~Andr{\'{e}}s
  Cisneros.
\newblock {LICHEM 1.1: Recent Improvements and New Capabilities}.
\newblock {\em Journal of Chemical Theory and Computation}, 15(5):3056--3065,
  may 2019.

\bibitem{Gordon2012}
Mark~S Gordon, Dmitri~G Fedorov, Spencer~R Pruitt, and Lyudmila~V Slipchenko.
\newblock {Fragmentation Methods: A Route to Accurate Calculations on Large
  Systems}.
\newblock {\em Chemical Reviews}, 112(1):632--672, jan 2012.

\bibitem{Govender2014}
Krishna Govender, Jiali Gao, and Kevin~J Naidoo.
\newblock {AM1/d-CB1: A Semiempirical Model for QM/MM Simulations of Chemical
  Glycobiology Systems}.
\newblock {\em Journal of chemical theory and computation}, 10:4694--4707,
  2014.

\bibitem{Gresh2007}
Nohad Gresh, G~Andr{\'{e}}s Cisneros, Thomas~A Darden, and Jean-Philip
  Piquemal.
\newblock {Anisotropic, Polarizable Molecular Mechanics Studies of Inter- and
  Intramolecular Interactions and Ligand-Macromolecule Complexes. A Bottom-Up
  Strategy}.
\newblock {\em Journal of chemical theory and computation}, 3(6):1960--1986,
  nov 2007.

\bibitem{Hagras2018}
Muhammad~A Hagras and William~J Glover.
\newblock {Polarizable Embedding for Excited-State Reactions: Dynamically
  Weighted Polarizable QM/MM}.
\newblock {\em Journal of Chemical Theory and Computation}, 14(4):2137--2144,
  apr 2018.

\bibitem{Halgren1999}
Thomas~A Halgren.
\newblock {MMFF VI. MMFF94s option for energy minimization studies}.
\newblock {\em Journal of Computational Chemistry}, 20(7):720--729, may 1999.

\bibitem{Henkelman2000}
Graeme Henkelman, Blas~P Uberuaga, and Hannes J{\'{o}}nsson.
\newblock {A climbing image nudged elastic band method for finding saddle
  points and minimum energy paths}.
\newblock {\em The Journal of Chemical Physics}, 113(22):9901--9904, 2000.

\bibitem{Holt2009}
Asbj{\o}rn Holt and Gunnar Karlstr{\"{o}}m.
\newblock {Improvement of the NEMO potential by inclusion of intramolecular
  polarization}.
\newblock {\em International Journal of Quantum Chemistry}, 109(6):1255--1266,
  may 2009.

\bibitem{Hornak2006}
Viktor Hornak, Robert Abel, Asim Okur, Bentley Strockbine, Adrian Roitberg, and
  Carlos Simmerling.
\newblock {Comparison of multiple Amber force fields and development of
  improved protein backbone parameters}.
\newblock {\em Proteins: Structure, Function, and Bioinformatics},
  65(3):712--725, nov 2006.

\bibitem{Hu2009}
Hao Hu and Weitao Yang.
\newblock {Development and application of ab initio QM/MM methods for
  mechanistic simulation of reactions in solution and in enzymes}.
\newblock {\em Theochem}, 898(1-3):17--30, mar 2009.

\bibitem{Im2001}
Wonpil Im, Simon Bern{\`{e}}che, and Benoıt Roux.
\newblock {Generalized solvent boundary potential for computer simulations}.
\newblock {\em The Journal of Chemical Physics}, 114(7):2924--2937, 2001.

\bibitem{Intharathep2006}
Pathumwadee Intharathep, Anan Tongraar, and Kritsana Sagarik.
\newblock {Ab initio QM/MM dynamics of H3O+ in water}.
\newblock {\em Journal of Computational Chemistry}, 27(14):1723--1732, 2006.

\bibitem{Jin2020}
Zhenming Jin, Xiaoyu Du, Yechun Xu, Yongqiang Deng, Meiqin Liu, Yao Zhao, Bing
  Zhang, Xiaofeng Li, Leike Zhang, Chao Peng, Yinkai Duan, Jing Yu, Lin Wang,
  Kailin Yang, Fengjiang Liu, Rendi Jiang, Xinglou Yang, Tian You, Xiaoce Liu,
  Xiuna Yang, Fang Bai, Hong Liu, Xiang Liu, Luke~W Guddat, Wenqing Xu, Gengfu
  Xiao, Chengfeng Qin, Zhengli Shi, Hualiang Jiang, Zihe Rao, and Haitao Yang.
\newblock {Structure of Mpro from SARS-CoV-2 and discovery of its inhibitors}.
\newblock {\em Nature}, 582(7811):289--293, 2020.

\bibitem{Jones32812020}
Leighton~O Jones, Mart{\'{i}}n~A Mosquera, George~C Schatz, and Mark~A Ratner.
\newblock {Embedding Methods for Quantum Chemistry: Applications from Materials
  to Life Sciences}.
\newblock {\em Journal of the American Chemical Society}, 142(7):3281--3295,
  feb 2020.

\bibitem{Jonsson1998}
H.~J{\'{o}}nsson, G.~Mills, and K.~W. Jacobsen.
\newblock {Nudged elastic band method for finding minimum energy paths of
  transitions}.
\newblock In {\em Classical and Quantum Dynamics in Condensed Phase
  Simulations}, pages 385--404. WORLD SCIENTIFIC, jun 1998.

\bibitem{Jorgensen1983}
William~L. Jorgensen, Jayaraman Chandrasekhar, Jeffry~D. Madura, Roger~W.
  Impey, and Michael~L. Klein.
\newblock {Comparison of simple potential functions for simulating liquid
  water}.
\newblock {\em The Journal of Chemical Physics}, 79(2):926--935, jul 1983.

\bibitem{Jorgensen1996}
William~L Jorgensen, David~S Maxwell, and Julian Tirado-Rives.
\newblock {Development and Testing of the OPLS All-Atom Force Field on
  Conformational Energetics and Properties of Organic Liquids}.
\newblock {\em Journal of the American Chemical Society}, 118(45):11225--11236,
  nov 1996.

\bibitem{Jung2005}
Yousung Jung, Alex Sodt, Peter M~W Gill, and Martin Head-Gordon.
\newblock {Auxiliary basis expansions for large-scale electronic structure
  calculations}.
\newblock {\em Proceedings of the National Academy of Sciences},
  102(19):6692--6697, 2005.

\bibitem{Kamerlin2010}
Shina C~L Kamerlin and Arieh Warshel.
\newblock {The EVB as a quantitative tool for formulating simulations and
  analyzing biological and chemical reactions}.
\newblock {\em Faraday Discuss.}, 145(0):71--106, 2010.

\bibitem{Komatsu2020}
Teruhisa~S. Komatsu, Yohei; Koyama, Noriaki; Okimoto, Gentaro; Morimoto,
  Yousuke; Ohno, and Makoto Taiji.
\newblock {COVID-19 related trajectory data of 10 microseconds all atom
  molecular dynamics simulation of SARS-CoV-2 dimeric main protease}.
\newblock {\em Mendeley Data}, 2020.

\bibitem{Kratz2016}
Eric~G Kratz, Robert~E Duke, and G~Andr{\'{e}}s Cisneros.
\newblock {Long-range electrostatic corrections in multipolar/polarizable QM/MM
  simulations}.
\newblock {\em Theoretical Chemistry Accounts}, 135(7):166, 2016.

\bibitem{Lichem2016}
Eric~G. Kratz, Alice~R. Walker, Louis Lagard{\`{e}}re, Filippo Lipparini,
  Jean-Philip Piquemal, and G.~{Andr{\'{e}}s Cisneros}.
\newblock {LICHEM: A QM/MM program for simulations with multipolar and
  polarizable force fields}.
\newblock {\em Journal of Computational Chemistry}, 37(11):1019--1029, apr
  2016.

\bibitem{TINKER-HP}
Louis Lagard{\`{e}}re, Luc-Henri Jolly, Filippo Lipparini, F{\'{e}}lix Aviat,
  Benjamin Stamm, Zhifeng~F Jing, Matthew Harger, Hedieh Torabifard,
  G~Andr{\'{e}}s Cisneros, Michael~J Schnieders, Nohad Gresh, Yvon Maday,
  Pengyu~Y Ren, Jay~W Ponder, and Jean-Philip Piquemal.
\newblock {Tinker-HP: a massively parallel molecular dynamics package for
  multiscale simulations of large complex systems with advanced point dipole
  polarizable force fields}.
\newblock {\em Chem. Sci.}, 9(4):956--972, 2018.

\bibitem{Lambros2020MBPolqmmm}
Eleftherios Lambros, Filippo Lipparini, G~Andres Cisneros, and Francesco
  Paesani.
\newblock {A Many-Body, Fully Polarizable Approach to QM/MM Simulations}.
\newblock {\em ChemRxiv}, sep 2020.

\bibitem{Li2015}
Pengfei Li, Lin~Frank Song, and Kenneth~M Merz.
\newblock {Parameterization of Highly Charged Metal Ions Using the 12-6-4
  LJ-Type Nonbonded Model in Explicit Water}.
\newblock {\em The Journal of Physical Chemistry B}, 119(3):883--895, jan 2015.

\bibitem{Lin2006}
Hai Lin and Donald~G Truhlar.
\newblock {QM/MM: what have we learned, where are we, and where do we go from
  here?}
\newblock {\em Theoretical Chemistry Accounts}, 117(2):185, 2006.

\bibitem{Lipparini2011}
Filippo Lipparini and Vincenzo Barone.
\newblock {Polarizable Force Fields and Polarizable Continuum Model: A
  Fluctuating Charges/PCM Approach. 1. Theory and Implementation}.
\newblock {\em Journal of Chemical Theory and Computation}, 7(11):3711--3724,
  nov 2011.

\bibitem{Loco2018}
Daniele Loco, Sandro Jurinovich, Lorenzo Cupellini, Maximilian F. S.~J. Menger,
  and Benedetta Mennucci.
\newblock {The modeling of the absorption lineshape for embedded molecules
  through a polarizable QM/MM approach}.
\newblock {\em Photochemical {\&} Photobiological Sciences}, 17(5):552--560,
  may 2018.

\bibitem{Loco2017}
Daniele Loco, Louis Lagard{\`{e}}re, Stefano Caprasecca, Filippo Lipparini,
  Benedetta Mennucci, and Jean-Philip Piquemal.
\newblock {Hybrid QM/MM Molecular Dynamics with AMOEBA Polarizable Embedding}.
\newblock {\em Journal of Chemical Theory and Computation}, 13(9):4025--4033,
  sep 2017.

\bibitem{Loco2019}
Daniele Loco, Louis Lagard{\`{e}}re, G{\'{e}}rardo~A. Cisneros, Giovanni
  Scalmani, Michael Frisch, Filippo Lipparini, Benedetta Mennucci, and
  Jean-Philip Piquemal.
\newblock {Towards large scale hybrid QM/MM dynamics of complex systems with
  advanced point dipole polarizable embeddings}.
\newblock {\em Chemical Science}, 10(30):7200--7211, jul 2019.

\bibitem{Loco2016}
Daniele Loco, {\'{E}}tienne Polack, Stefano Caprasecca, Louis Lagard{\`{e}}re,
  Filippo Lipparini, Jean-Philip Piquemal, and Benedetta Mennucci.
\newblock {A QM/MM Approach Using the AMOEBA Polarizable Embedding: From Ground
  State Energies to Electronic Excitations}.
\newblock {\em Journal of Chemical Theory and Computation}, 12(8):3654--3661,
  aug 2016.

\bibitem{Lopes2009b}
Pedro E~M Lopes, Guillaume Lamoureux, and Alexander~D {Mackerell Jr.}
\newblock {Polarizable empirical force field for nitrogen-containing
  heteroaromatic compounds based on the classical Drude oscillator}.
\newblock {\em Journal of Computational Chemistry}, 30(12):1821--1838, sep
  2009.

\bibitem{Lopes2009}
Pedro E~M Lopes, Benoit Roux, and Alexander~D MacKerell.
\newblock {Molecular modeling and dynamics studies with explicit inclusion of
  electronic polarizability: theory and applications}.
\newblock {\em Theoretical Chemistry Accounts}, 124(1):11--28, 2009.

\bibitem{Magalhaes2020}
Rita~P Magalh{\~{a}}es, Henriques~S Fernandes, and S{\'{e}}rgio~F Sousa.
\newblock {Modelling Enzymatic Mechanisms with QM/MM Approaches: Current Status
  and Future Challenges}.
\newblock {\em Israel Journal of Chemistry}, 60(7):655--666, 2020.

\bibitem{Mao2017}
Yuezhi Mao, Yihan Shao, Jacek Dziedzic, Chris-Kriton Skylaris, Teresa
  Head-Gordon, and Martin Head-Gordon.
\newblock {Performance of the AMOEBA Water Model in the Vicinity of QM Solutes:
  A Diagnosis Using Energy Decomposition Analysis.}
\newblock {\em Journal of chemical theory and computation}, 13(5):1963--1979,
  may 2017.

\bibitem{McCann2013}
Billy~W McCann and Orlando Acevedo.
\newblock {Pairwise Alternatives to Ewald Summation for Calculating Long-Range
  Electrostatics in Ionic Liquids.}
\newblock {\em Journal of chemical theory and computation}, 9(2):944--950, feb
  2013.

\bibitem{Monard1996}
G{\'{e}}rald Monard, Michel Loos, Vincent Th{\'{e}}ry, Kristofor Baka, and
  Jean-Louis Rivail.
\newblock {Hybrid classical quantum force field for modeling very large
  molecules}.
\newblock {\em International Journal of Quantum Chemistry}, 58(2):153--159, jan
  1996.

\bibitem{Monari2013}
Antonio Monari, Jean-Louis Rivail, and Xavier Assfeld.
\newblock {Theoretical Modeling of Large Molecular Systems. Advances in the
  Local Self Consistent Field Method for Mixed Quantum Mechanics/Molecular
  Mechanics Calculations}.
\newblock {\em Accounts of Chemical Research}, 46(2):596--603, feb 2013.

\bibitem{Munoz-Munoz2015}
Y~Mauricio Mu{\~{n}}oz-Mu{\~{n}}oz, Gabriela Guevara-Carrion, Mario
  Llano-Restrepo, and Jadran Vrabec.
\newblock {Lennard-Jones force field parameters for cyclic alkanes from
  cyclopropane to cyclohexane}.
\newblock {\em Fluid Phase Equilibria}, 404:150--160, 2015.

\bibitem{Nam2005}
Kwangho Nam, Jiali Gao, and Darrin~M York.
\newblock {An Efficient Linear-Scaling Ewald Method for Long-Range
  Electrostatic Interactions in Combined QM/MM Calculations}.
\newblock {\em Journal of Chemical Theory and Computation}, 1(1):2--13, jan
  2005.

\bibitem{Notman2013}
Rebecca Notman and Jamshed Anwar.
\newblock {Breaching the skin barrier — Insights from molecular simulation of
  model membranes}.
\newblock {\em Advanced Drug Delivery Reviews}, 65(2):237--250, 2013.

\bibitem{Ojeda-May2017}
Pedro Ojeda-May and Kwangho Nam.
\newblock {Acceleration of Semiempirical QM/MM Methods through Message Passage
  Interface (MPI), Hybrid MPI/Open Multiprocessing, and Self-Consistent Field
  Accelerator Implementations}.
\newblock {\em Journal of Chemical Theory and Computation}, 13(8):3525--3536,
  aug 2017.

\bibitem{Pan2018}
Xiaoliang Pan, Edina Rosta, and Yihan Shao.
\newblock {Representation of the QM Subsystem for Long-Range Electrostatic
  Interaction in Non-Periodic Ab Initio QM/MM Calculations}.
\newblock {\em Molecules}, 23(10):2500, 2018.

\bibitem{Parks2008}
Jerry~M. Parks, Hao Hu, Aron~J. Cohen, and Weitao Yang.
\newblock {A pseudobond parametrization for improved electrostatics in quantum
  mechanical/molecular mechanical simulations of enzymes}.
\newblock {\em The Journal of Chemical Physics}, 129(15):154106, oct 2008.

\bibitem{psi4}
Robert~M Parrish, Lori~A Burns, Daniel G~A Smith, Andrew~C Simmonett, A~Eugene
  DePrince, Edward~G Hohenstein, Uğur Bozkaya, Alexander~Yu. Sokolov, Roberto
  {Di Remigio}, Ryan~M Richard, J{\'{e}}r{\^{o}}me~F Gonthier, Andrew~M James,
  Harley~R McAlexander, Ashutosh Kumar, Masaaki Saitow, Xiao Wang, Benjamin~P
  Pritchard, Prakash Verma, Henry~F Schaefer, Konrad Patkowski, Rollin~A King,
  Edward~F Valeev, Francesco~A Evangelista, Justin~M Turney, T~Daniel Crawford,
  and C~David Sherrill.
\newblock {Psi4 1.1: An Open-Source Electronic Structure Program Emphasizing
  Automation, Advanced Libraries, and Interoperability}.
\newblock {\em Journal of Chemical Theory and Computation}, 13(7):3185--3197,
  jul 2017.

\bibitem{Piquemal2006}
Jean-Philip Piquemal, G~Andr{\'{e}}s Cisneros, Peter Reinhardt, Nohad Gresh,
  and Thomas~A Darden.
\newblock {Towards a force field based on density fitting}.
\newblock {\em The Journal of Chemical Physics}, 124(10):104101, 2006.

\bibitem{LAMMPS}
Steve Plimpton.
\newblock {Fast Parallel Algorithms for Short-Range Molecular Dynamics}.
\newblock {\em Journal of Computational Physics}, 117(1):1--19, 1995.

\bibitem{Ponder2010}
Jay~W Ponder, Chuanjie Wu, Pengyu Ren, Vijay~S Pande, John~D Chodera, Michael~J
  Schnieders, Imran Haque, David~L Mobley, Daniel~S Lambrecht, Robert~A
  DiStasio, Martin Head-Gordon, Gary N~I Clark, Margaret~E Johnson, and Teresa
  Head-Gordon.
\newblock {Current Status of the AMOEBA Polarizable Force Field}.
\newblock {\em The Journal of Physical Chemistry B}, 114(8):2549--2564, mar
  2010.

\bibitem{TINKER}
Joshua~A Rackers, Zhi Wang, Chao Lu, Marie~L Laury, Louis Lagard{\`{e}}re,
  Michael~J Schnieders, Jean-Philip Piquemal, Pengyu Ren, and Jay~W Ponder.
\newblock {Tinker 8: Software Tools for Molecular Design}.
\newblock {\em Journal of chemical theory and computation}, 14(10):5273--5289,
  oct 2018.

\bibitem{Rappe1992}
A~K Rappe, C~J Casewit, K~S Colwell, W~A Goddard, and W~M Skiff.
\newblock {UFF, a full periodic table force field for molecular mechanics and
  molecular dynamics simulations}.
\newblock {\em Journal of the American Chemical Society}, 114(25):10024--10035,
  dec 1992.

\bibitem{Reddy1945042016}
Sandeep~K Reddy, Shelby~C Straight, Pushp Bajaj, C~{Huy Pham}, Marc Riera,
  Daniel~R Moberg, Miguel~A Morales, Chris Knight, Andreas~W G{\"{o}}tz, and
  Francesco Paesani.
\newblock {On the accuracy of the MB-pol many-body potential for water:
  Interaction energies, vibrational frequencies, and classical thermodynamic
  and dynamical properties from clusters to liquid water and ice.}
\newblock {\em The Journal of chemical physics}, 145(19):194504, nov 2016.

\bibitem{Riccardi2004}
Demian Riccardi, Guohui Li, and Qiang Cui.
\newblock {Importance of van der Waals Interactions in QM/MM Simulations}.
\newblock {\em The Journal of Physical Chemistry B}, 108(20):6467--6478, may
  2004.

\bibitem{Sarkar2011}
Anirban Sarkar, Sudipta~Raha Roy, Naisargee Parikh, and Asit~K. Chakraborti.
\newblock {Nonsolvent Application of Ionic Liquids: Organo-Catalysis by
  1-Alkyl-3-methylimidazolium Cation Based Room-Temperature Ionic Liquids for
  Chemoselective {\textless}i{\textgreater}N{\textless}/i{\textgreater} -
  {\textless}i{\textgreater}tert{\textless}/i{\textgreater}
  -Butyloxycarbonylation of Amines and the Influence of the C-2 Hydrogen on
  Catalytic Efficiency}.
\newblock {\em The Journal of Organic Chemistry}, 76(17):7132--7140, sep 2011.

\bibitem{Scott1999}
Walter R~P Scott, Philippe~H H{\"{u}}nenberger, Ilario~G Tironi, Alan~E Mark,
  Salomon~R Billeter, Jens Fennen, Andrew~E Torda, Thomas Huber, Peter
  Kr{\"{u}}ger, and Wilfred~F van Gunsteren.
\newblock {The GROMOS Biomolecular Simulation Program Package}.
\newblock {\em The Journal of Physical Chemistry A}, 103(19):3596--3607, may
  1999.

\bibitem{Senn2009}
Hans~Martin Senn and Walter Thiel.
\newblock {QM/MM Methods for Biomolecular Systems}.
\newblock {\em Angewandte Chemie International Edition}, 48(7):1198--1229, feb
  2009.

\bibitem{Singh1986}
U.~Chandra Singh and Peter~A. Kollman.
\newblock {A combined ab initio quantum mechanical and molecular mechanical
  method for carrying out simulations on complex molecular systems:
  Applications to the CH3Cl + Cl- exchange reaction and gas phase protonation
  of polyethers}.
\newblock {\em Journal of Computational Chemistry}, 7(6):718--730, dec 1986.

\bibitem{Strajbl2002}
Marek {\v{S}}trajbl, Gongyi Hong, and Arieh Warshel.
\newblock {Ab Initio QM/MM Simulation with Proper Sampling: “First
  Principle” Calculations of the Free Energy of the Autodissociation of Water
  in Aqueous Solution}.
\newblock {\em The Journal of Physical Chemistry B}, 106(51):13333--13343, dec
  2002.

\bibitem{Thery1994}
Vincent Th{\'{e}}ry, Daniel Rinaldi, Jean-Louis Rivail, Bernard Maigret, and
  Gy{\"{o}}rgy~G. Ferenczy.
\newblock {Quantum mechanical computations on very large molecular systems: The
  local self-consistent field method}.
\newblock {\em Journal of Computational Chemistry}, 15(3):269--282, mar 1994.

\bibitem{Tschumper02690}
Gregory~S Tschumper, Matthew~L Leininger, Brian~C Hoffman, Edward~F Valeev,
  Henry~F Schaefer, and Martin Quack.
\newblock {Anchoring the water dimer potential energy surface with explicitly
  correlated computations and focal point analyses}.
\newblock {\em The Journal of Chemical Physics}, 116(2):690--701, dec 2002.

\bibitem{NWChem}
M~Valiev, E~J Bylaska, N~Govind, K~Kowalski, T~P Straatsma, H~J~J {Van Dam},
  D~Wang, J~Nieplocha, E~Apra, T~L Windus, and W~A de~Jong.
\newblock {NWChem: A comprehensive and scalable open-source solution for large
  scale molecular simulations}.
\newblock {\em Computer Physics Communications}, 181(9):1477--1489, 2010.

\bibitem{VanderKamp2013}
Marc~W van~der Kamp and Adrian~J Mulholland.
\newblock {Combined Quantum Mechanics/Molecular Mechanics (QM/MM) Methods in
  Computational Enzymology}.
\newblock {\em Biochemistry}, 52(16):2708--2728, apr 2013.

\bibitem{VanVleet2016}
Mary~J {Van Vleet}, Alston~J Misquitta, Anthony~J Stone, and J~R Schmidt.
\newblock {Beyond Born–Mayer: Improved Models for Short-Range Repulsion in ab
  Initio Force Fields}.
\newblock {\em Journal of Chemical Theory and Computation}, 12(8):3851--3870,
  aug 2016.

\bibitem{Vazquez-Montelongo2018}
Erik V{\'{a}}zquez-Montelongo, Jos{\'{e}} V{\'{a}}zquez-Cervantes, and
  G.~Cisneros.
\newblock {Polarizable ab initio QM/MM Study of the Reaction Mechanism of
  N-tert-Butyloxycarbonylation of Aniline in [EMIm][BF4]}.
\newblock {\em Molecules}, 23(11):2830, oct 2018.

\bibitem{ViquezRojas15492020}
Claudia~I {Viquez Rojas} and Lyudmila~V Slipchenko.
\newblock {Exchange-repulsion in QM/EFP excitation energies – beyond
  polarizable embedding}.
\newblock {\em Journal of Chemical Theory and Computation}, aug 2020.

\bibitem{Walker2008}
Ross~C Walker, Michael~F Crowley, and David~A Case.
\newblock {The implementation of a fast and accurate QM/MM potential method in
  Amber.}
\newblock {\em Journal of computational chemistry}, 29(7):1019--1031, may 2008.

\bibitem{Warshel1976}
A~Warshel and M~Levitt.
\newblock {Theoretical studies of enzymic reactions: Dielectric, electrostatic
  and steric stabilization of the carbonium ion in the reaction of lysozyme}.
\newblock {\em Journal of Molecular Biology}, 103(2):227--249, 1976.

\bibitem{Warshel1980}
Arieh Warshel and Robert~M Weiss.
\newblock {An empirical valence bond approach for comparing reactions in
  solutions and in enzymes}.
\newblock {\em Journal of the American Chemical Society}, 102(20):6218--6226,
  sep 1980.

\bibitem{Watanabe2019}
Hiroshi~C Watanabe and Qiang Cui.
\newblock {Quantitative Analysis of QM/MM Boundary Artifacts and Correction in
  Adaptive QM/MM Simulations}.
\newblock {\em Journal of Chemical Theory and Computation}, 15(7):3917--3928,
  jul 2019.

\bibitem{Wheatly-Price-1990}
Richard~J Wheatley and Sarah~L Price.
\newblock {An overlap model for estimating the anisotropy of repulsion}.
\newblock {\em Molecular Physics}, 69(3):507--533, feb 1990.

\bibitem{Xiao2018}
Gaobo Xiao, Mingjun Ren, and Haibo Hong.
\newblock 50 million atoms scale molecular dynamics modelling on a single
  consumer graphics card.
\newblock {\em Advances in Engineering Software}, 124:66--72, 2018.

\bibitem{Xie2004}
Li~Xie, Haiyan Liu, and Weitao Yang.
\newblock {Adapting the nudged elastic band method for determining
  minimum-energy paths of chemical reactions in enzymes}.
\newblock {\em The Journal of Chemical Physics}, 120(17):8039--8052, apr 2004.

\bibitem{Zhang2005}
Yingkai Zhang.
\newblock {Improved pseudobonds for combined {\textless}i{\textgreater}ab
  initio{\textless}/i{\textgreater} quantum mechanical/molecular mechanical
  methods}.
\newblock {\em The Journal of Chemical Physics}, 122(2):024114, jan 2005.

\bibitem{Zhang1999}
Yingkai Zhang, Tai-Sung Lee, and Weitao Yang.
\newblock {A pseudobond approach to combining quantum mechanical and molecular
  mechanical methods}.
\newblock {\em The Journal of Chemical Physics}, 110(1):46--54, jan 1999.

\end{thebibliography}

\end{document}